\documentclass[prd,showpacs,amsmath,amssymb,nofootinbib,superscriptaddress,notitlepage,preprintnumbers]{revtex4-1}
\usepackage[usenames,dvipsnames,svgnames,table]{xcolor} 
\usepackage{array,mathtools,amssymb,booktabs,multirow,diagbox,hhline,natbib}
\pdfoutput=1

\usepackage{graphicx,graphics,color}
\usepackage{dcolumn}
\usepackage{bm}
\usepackage{bbold}
\usepackage{hyperref}
\hypersetup{ 
    pdfnewwindow=true,      
    colorlinks=true,       
    allcolors=SeaGreen
} 
\usepackage{amsmath}
\usepackage{xcolor}
\usepackage{tikz}
\usepackage{tikz-feynman}
\usepackage{simpler-wick}
\usepackage{subfigure}

\allowdisplaybreaks

\newcommand{\Od}{{\cal O}}
\newcommand{\im}{\mbox{Im}\,}
\newcommand{\re}{\mbox{Re}\,}
\newcommand{\sgn}{\mbox{sgn}}

\newcommand{\intT}{\int_0^{1/T} d\tau \int d^3 \vec{x}}

\newcommand{\modQ}{\vert \vec{Q} \vert}
\newcommand{\tsum}{T \sum_{n=-\infty}^{\infty}}
\newcommand{\sumint}{\sum \! \! \! \! \! \! \! \!   \int}
\newcommand{\mean}[1]{\left\langle{#1}\right\rangle}

\newcommand{\conds}{\langle \bar s s \rangle}
\newcommand{\condl}{\mean{\bar q q}_l}

\newcommand{\Sr}{\textrm{\bf S}}
\newcommand{\Tr}{\textrm{\bf T}}
\newcommand{\Ur}{\textrm{\bf U}}
\newcommand{\IR}{{\Bbb R}}

\newcommand{\gsim}{\raise.3ex\hbox{$>$\kern-.75em\lower1ex\hbox{$\sim$}}}

\begin{document}

\preprint{IPARCOS-UCM-23-076}

\title{The pion-kaon scattering amplitude  and the $K_0^* (700)$ and $K^*(892)$ resonances at finite temperature}
\author{A. G\'omez Nicola}
\email{gomez@ucm.es}
\affiliation{Departamento de F\'isica Te\'orica and
IPARCOS, Universidad Complutense de Madrid, Plaza de las Ciencias 1, 28040 Madrid, Spain}
\author{J. Ruiz de Elvira}
\email{jacobore@ucm.es}
\affiliation{Departamento de F\'isica Te\'orica and
IPARCOS, Universidad Complutense de Madrid, Plaza de las Ciencias 1, 28040 Madrid, Spain}
\author{A. Vioque-Rodr\'iguez}
\email{avioque@ucm.es}
\affiliation{Departamento de F\'isica Te\'orica and
IPARCOS, Universidad Complutense de Madrid, Plaza de las Ciencias 1, 28040 Madrid, Spain}

\begin{abstract}
 We perform a complete calculation of the pion-kaon scattering amplitude in Chiral Perturbation Theory at finite temperature, paying particular attention to the analytic structure of the amplitude and the main differences with respect to the zero temperature case. We also extend the Inverse Amplitude Method at finite temperature for unequal-mass scattering processes, which allows us to unitarize the amplitude and obtain the thermal evolution of the $K_0^* (700)$ and $K^*(892)$ pole parameters. As a direct application of our analysis, we show that the thermal evolution of the $K_0^* (700)$ resonance is crucial to explain the behavior of the scalar susceptibility for isospin $I=1/2$, which in turn, is directly connected with chiral and $U(1)_A$ restoration properties of the QCD phase diagram. 
 \end{abstract}

 \maketitle

\section{Introduction}

Over recent years, hadronic matter under conditions of temperature and chemical potentials relevant to the QCD phase diagram has been the object of intense study.  
Theoretical tools, based mostly on effective field theories~\cite{Pisarski:1983ms,Hatsuda:1985eb,Bernard:1987im,Gerber:1988tt,Venugopalan:1992hy,Schenk:1993ru,Bochkarev:1995gi,Dobado:1998tv,Rapp:1999ej,Ayala:2000px,GomezNicola:2002tn,Dobado:2002xf,Karsch:2003vd,Huovinen:2009yb,FernandezFraile:2009mi,Costa:2010zw,Jankowski:2012ms,GomezNicola:2010tb,Nicola:2013vma,GomezNicola:2016ssy,Ishii:2016dln,GomezNicola:2017bhm,Nicola:2018vug,GomezNicola:2019myi,Nicola:2020iyl,Nicola:2020smo}, the rapid development of lattice simulations~\cite{Aoki:2009sc,Bazavov:2011nk,Buchoff:2013nra,Cossu:2013uua,Brandt:2016daq,Tomiya:2016jwr,Bazavov:2018mes,Ding:2019prx,Ratti:2018ksb,Bazavov:2019lgz} and even  experimental information within the Beam Energy Scan program in heavy-ion collisions~\cite{Adamczyk:2017iwn,Andronic:2017pug} have boosted the activity and knowledge within this field. 

The emerging consistent picture is that the QCD transition of deconfinement and chiral symmetry restoration takes place in the plane of temperature $T$ and baryon chemical potential $\mu_B$ as a smooth crossover at low $\mu_B$, which would turn into a first-order transition at the QCD critical point. The existence and properties of the latter constitute one of the open problems in the field~\cite{Bazavov:2018mes,Ratti:2018ksb,Bazavov:2019lgz}, together with the nature of the transition, and its connection with $U(1)_A$ restoration, which is studied mostly through the degeneration of susceptibilities and screening masses in different channels~\cite{Pisarski:1983ms,Ishii:2016dln,GomezNicola:2017bhm,Nicola:2018vug,GomezNicola:2019myi,Nicola:2020iyl,Nicola:2020smo,Buchoff:2013nra,Cossu:2013uua,Brandt:2016daq,Dick:2015twa,Tomiya:2016jwr,Shuryak:1993ee,Kapusta:1995ww,Cohen:1996ng,Lee:1996zy,Meggiolaro:2013swa,Pelissetto:2013hqa}. At $\mu_B=0$, the crossover transition takes place at a critical temperature $T_c\simeq 155$ MeV, which goes down to $T_c^0\simeq$ 129 MeV in the light chiral limit $m_{u,d}\rightarrow 0^+$~\cite{Ding:2019prx}. In that case, the transition is possibly of second order, although it could be of first order if the $U(1)_A$ symmetry is sufficiently weak near $T_c$~\cite{Pisarski:1983ms,Shuryak:1993ee,Pelissetto:2013hqa}.

A relevant part of this program has been to include the effect of interactions among the thermal bath components and understand their role in the phase diagram, especially regarding chiral symmetry restoration. Actually, for certain  observables, it turns out that including properly the thermal (or in-medium) modifications of their spectral properties, such as the mass and width of the resonances that can be created and decay in the thermal bath, is more relevant than including heavier states, as customarily done in approaches based on the Hadron Resonance Gas (HRG) model~\cite{Karsch:2003vd,Huovinen:2009yb,Jankowski:2012ms}. A very significant example is the light scalar susceptibility, i.e., the correlator of the quark condensate at vanishing momentum. This is one of the key observables signaling chiral symmetry restoration since it  develops a peak at the transition temperature in the crossover regime that should get stronger as the light $u,d$ quark masses decrease, becoming a divergence in the light chiral limit if the transition is of second order~\cite{Smilga:1995qf,Aoki:2009sc,Bazavov:2011nk,Bazavov:2018mes,Ding:2019prx,Ratti:2018ksb,Bazavov:2019lgz}. In fact, in recent years it has been shown that saturating the scalar susceptibility with the lightest meson state with its quantum numbers of isospin and total angular momentum $I=J=0$, i.e., the $f_0(500)$ thermal resonance, yields the expected peaked profile in accordance with lattice results and improves the description of the scalar susceptibility around the transition over the HRG~\cite{Nicola:2013vma,Ferreres-Sole:2018djq}. 
The spectral properties of the $f_0(500)$  resonance have been obtained from the second Riemann sheet pole of the $I=J=0$ partial wave of the $\pi\pi$ scattering amplitude at finite temperature in unitarized Chiral Perturbation Theory (ChPT)~\cite{GomezNicola:2002tn,Dobado:2002xf},  which has proven to be a quite successful scheme to describe light meson spectroscopy and thermal properties~\cite{Gerber:1988tt,Schenk:1993ru,Dobado:1998tv,FernandezFraile:2009mi,Nicola:2013vma,GomezNicola:2017bhm,Nicola:2018vug}. While ChPT provides the most general low-energy Lagrangian and a consistent perturbative scheme compatible with the QCD symmetries~\cite{Gasser:1983yg,Gasser:1984gg,Gerber:1988tt}, unitarization methods allow one to extend the ChPT applicability range and generate dynamically the expected resonance spectrum~\cite{Dobado:1996ps,Oller:1997ti,Oller:1998hw,GomezNicola:2001as,RuizdeElvira:2010cs,Guo:2012ym,Pelaez:2015qba,RuizdeElvira:2018hsv,Pelaez:2021dak}. Actually, the same scheme yields in the $I=J=1$ channel the thermal modifications of the mass and width of the $\rho (770)$ resonance at finite temperature~\cite{GomezNicola:2002tn,Dobado:2002xf,GomezNicola:2004gg} in agreement with the  in-medium broadening of that resonance expected from other theoretical analyses and from the experimental dilepton spectrum \cite{Rapp:1999ej,Song:1996dg,Rapp:2014hha,ALICE:2018ael}. In addition, the study of the thermal spectral functions of the $\rho (770)$ and the $a_1(1260)$ mesons shows that those states become degenerate at the chiral transition, as it should be expected~\cite{Rapp:1999ej,Jung:2016yxl}. 

In the present work, we extend the previous program to study pion-kaon scattering. Namely, we will compute the $\pi K$ elastic scattering amplitude at finite temperature within ChPT, as well as study its unitarization and the generation of the thermal $K_0^* (700)$ and $K^*(890)$ resonances. Updated analyses and reviews about these resonances at $T=0$ can be found, for e.g., in~\cite{Pelaez:2021dak,Pelaez:2016klv,Pelaez:2020uiw,Pelaez:2020gnd}, where their present theoretical and experimental status, as well as precise determinations of their spectral properties, are studied. 

The interest in the $K_0^* (700)$ resonance at finite temperature has increased lately. For instance, in~\cite{Azizi:2019kzj}, its thermal properties are studied within thermal sum rules, and in~\cite{Giacosa:2018vbw}, using virial expansion methods. Nevertheless, these two references do not take into account the thermal modifications of the $\pi K$ amplitude at finite temperature.  Our analysis also has direct implications for the chiral transition and its nature since, in a recent work~\cite{GomezNicola:2020qxo}, it has been proved using Ward Identities (WI) that the $I=1/2$ scalar susceptibility should also have a peak above $T_c$. This peak indicates the onset of $U_A(1)$ restoration via the degeneration of the scalar and pseudoscalar channels, whose lightest states are the $K_0^* (700)$ and the kaon, respectively. In addition, in~\cite{GomezNicola:2020qxo}, it has been shown that such a peak can be reproduced by saturating the $I=1/2$ scalar susceptibility with the $K_0^* (700)$ thermal pole, which, in turn, is generated via the unitarization of a simplified $\pi K$ thermal amplitude. While at $T=0$ such amplitude corresponds to the full ChPT prediction, at finite temperature it only includes the $S$-channel contribution responsible for thermal unitarity, along the lines discussed in~\cite{Gao:2019idb}. 

Therefore, the purpose of the present work is to provide the full calculation of the $\pi K$ scattering amplitude in ChPT at finite temperature and analyze its main phenomenological consequences for the topics discussed above related to the QCD phase diagram.  The main advantages and novelties of our analysis are the following: 

\begin{enumerate}

\item  By construction, ChPT includes the correct thermal dependence of any Goldstone-boson scattering amplitude at low temperatures, not only the effects related to thermal unitarity. In particular, as we will see in detail, the analytical structure of a thermal amplitude gets much more complicated due to the loop integrals involved. We will then incorporate all effects properly, of which those weighted by Bose-Einstein distribution functions evaluated at the pion mass are expected to have a significative effect near $T_c$. In particular, we will include thermal tadpoles, which were neglected in~\cite{GomezNicola:2020qxo}, and only partially considered in~\cite{Gao:2019idb} as corrections to thermal masses. 

\item The pion-kaon  amplitude is renormalized consistently within the standard ChPT dimensional regularization scheme, where the low-energy constants (LECs) absorb ultraviolet divergences at $T=0$. Therefore, we will be able to use recent LECs determinations when performing our numerical analysis.

\item The pion and kaon mass dependence of the amplitude is under control within ChPT. This will be particularly useful when discussing the light chiral limit, which is of great relevance for chiral and $U(1)_A$ restoration, as well as the behavior towards the $SU(3)$ limit of pion-kaon degeneration\footnote{Note that studying the exact $SU(3)$ limit would require a coupled-channel analysis since, in this case, the $\eta K$ and $\pi K$ thresholds coincide. Instead, we will study only the behavior towards $SU(3)$ degeneration, limiting ourselves to kaon masses for which an elastic approximation still makes sense.}.  

\item The complete $\pi K$ perturbative amplitude and its unitarization at finite temperature will provide a rigorous check of the consistency and robustness of  previous approaches for the $I=1/2,J=0$ channel regarding the thermal behavior of the $K_0^* (700)$ pole, as well as its connection with chiral restoration. 

\item We obtain in turn the $I=1/2, J=1$ vector partial wave at finite temperature. Thus, we also study the thermal properties of the $K^*(892)$ pole, which has not been studied before in this context. Note that while scalars meson can be reproduced reasonably well within the type of unitarization methods used in~\cite{Gao:2019idb,GomezNicola:2020qxo}, vector mesons require an accurate fourth-order ChPT description. 
The $K^*(892)$ meson can be indeed produced in heavy-ion collisions~\cite{STAR:2002npn}, and recent estimations predict small in-medium modifications (temperature and baryon chemical potential) of its spectral properties compared with the $\rho (770)$~\cite{Reichert:2022uha}, consistently with not observing a significative reduction in $K^*(892)$ production. 
\end{enumerate}

This article is organized as follows. In section~\ref{sec:amp} we present the general features of the $\pi K$ thermal amplitude and calculate it in ChPT. In section \ref{sec:anther} we discuss the modifications of the amplitude analytical structure induced by finite-temperature corrections. These corrections are quite different from those in $\pi\pi$ scattering discussed in \cite{GomezNicola:2002tn}. The fact that pion-kaon scattering is an unequal-mass process gives rise to the appearance of the so-called thermal Landau cuts, related to scattering processes taking place in the thermal bath. Thus, we obtain here the generalization for unequal masses of the thermal unitarity relation for the perturbative amplitude obtained in~\cite{GomezNicola:2002tn}.  The temperature modification of partial waves and scattering lengths in ChPT is obtained in  section~\ref{sec:numchpt} as a direct phenomenological consequence of our study.  In section~\ref{sec:unit} we construct a unitarized thermal amplitude from the perturbative ChPT one, following the guidelines of the Inverse Amplitude Method (IAM) at finite temperature. From the unitarized amplitude, we will calculate the temperature dependence of the $I=1/2, J=0$ pole parameters, mass, width, and residue, and compare them with previous results. This pole is used in section~\ref{sec:sus} to saturate the scalar $I=1/2$ susceptibility. We will see that the expected behavior is reproduced and we will provide a comparison with previous analysis. Special attention is paid to the effect of the LECs uncertainties in our results. As commented, our analysis is also suitable to reproduce the $K^*(892)$ thermal properties, which is carried out in section~\ref{sec:unit}.
Finally, we have moved to the Appendices all the technical details regarding kinematics, thermal loop integrals, and other issues, as well as the explicit expression for the thermal amplitude.

\section{The $\pi K$ scattering amplitude at finite temperature in ChPT}
\label{sec:amp} 

For the definition of the scattering amplitude at finite temperature, we follow the standard prescription and assume that the temperature dependence arises in the loops from the Thermal Field Theory Feynman rules~\cite{galekapustabook}, i.e., in the four-point Green function connected with the ${\cal T}$-matrix elements through the LSZ reduction formula.  Within this approach, the $\pi\pi$ scattering lengths~\cite{Quack:1994vc,Loewe:2008kh} and the $I=J=1$ partial wave~\cite{He:1997gn} have been computed at finite temperature in the Nambu-Jona-Lasinio (NJL) model. The scattering lengths have also been calculated within the Linear Sigma Model (LSM)~\cite{Loewe:2008ui} and in ChPT~\cite{Kaiser:1999mt}, while, as mentioned above, the full ChPT $\pi\pi$ scattering finite-temperature amplitude  was obtained in~\cite{GomezNicola:2002tn}.  As for the $\pi K$ elastic thermal amplitude, the only available analyses to our knowledge are those in~\cite{Gao:2019idb,GomezNicola:2020qxo}, which provide a partial calculation of the amplitude since the sole one-loop effects included are those related to thermal unitarity (see below).

\begin{figure}[h]
 \centering
\includegraphics[width=12cm]{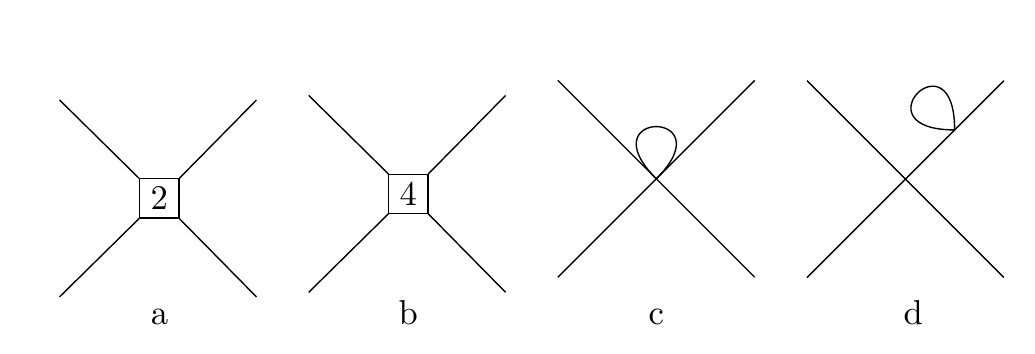}
\includegraphics[width=12cm]{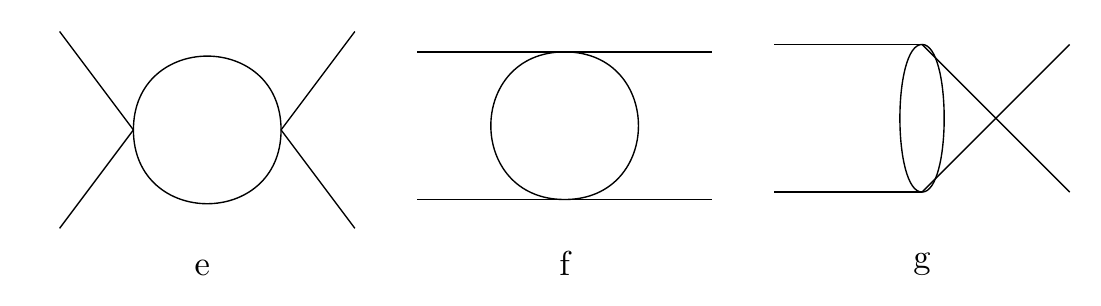}
\caption{Feynman diagrams corresponding to the  $K\pi \rightarrow K\pi$ scattering amplitude  up to fourth order in ChPT.}
\label{fig:diags}
\end{figure}

 The ChPT framework guarantees that the low-energy amplitude includes all possible terms compatible with the QCD symmetries and, especially, the chiral symmetry breaking pattern within a consistent low-energy chiral power counting, renormalizable order by order. Here, we are interested in the elastic $K\,\pi \rightarrow K\,\pi $ scattering amplitude. The relevant type of Feynman diagrams are shown in Fig.~\ref{fig:diags}. We are considering the amplitude up to $\Od(p^4)$ in the chiral power counting, so that the vertices and propagators entering those diagrams (the same as for $T=0$) are calculated from the ChPT $SU(3)$ Lagrangians ${\cal L}_2$  and ${\cal L}_4$ given in ~\cite{Gasser:1984gg}, with ${\cal L}_n$ denoting the $\Od(p^n)$ Lagrangian and $p$ standing for a generic low-energy scale such as meson momenta, masses or temperature. Note that this implies that we neglect any multipion contributions, which are suppressed at low energies by their multibody phase space and have not been observed experimentally below 1 GeV at $T=0$.

 The general structure of the amplitude is as follows. First, as for $T=0$, one has to consider the contributions from the  $S,\,T\,,U-$ channels corresponding to the different ways of pairing incoming and outgoing external momenta  in the reaction $p_K\,p_\pi  \rightarrow p'_K\,p'_\pi $ as shown in Fig.~\ref{fig:diags}.  For every one of those channels and for the sum of them,  the amplitude at finite temperature $T$, e.g., for the process $K^+\pi^+\rightarrow K^+\pi^+$, can be written according to the  general structure
 
 \begin{equation}
 {\cal T} (\Sr,\Tr,\Ur;T)={\cal T}_2 (s,t,u)+ {\cal T}_4^{tree}(s,t,u)+ {\cal T}_4^F (\Sr,\Tr,\Ur;T)  + {\cal T}_4^J (\Sr,\Tr,\Ur;T),
 \end{equation}
 where
 $\Sr=p_K+p_\pi$, $\Tr=p_K-p'_K$, $\Ur=p_K-p'_\pi$, being  $s=\Sr^2$, $t=\Tr^2$, and $u=\Ur^2$ the usual Mandelstam variables.  The different contributions to the amplitude read as follows: ${\cal T}_2 (s,t,u)$ is the tree-level $\Od(p^2)$ contribution to the amplitude coming from the ${\cal L}_2$ Lagrangian (diagram (a) in Fig.~\ref{fig:diags}), ${\cal T}_4^{tree}$ is the $\Od(p^4)$ tree-level ${\cal L}_4$ contribution showed in diagram (b) in Fig.~\ref{fig:diags}, and ${\cal T}_4^F$, ${\cal T}_4^J$ are the one-loop contributions, from which the temperature dependence arises,   standing for the tadpole-like and $J$-like thermal integrals (we follow the same convention as in~\cite{GomezNicola:2002tn}), defined by the loop functions
  \begin{equation}
 F_{\beta a} (T)=\sumint_q \frac{1}{q^2-M_a^2},
 \label{tadpole}
 \end{equation}
 and 
 \begin{equation}
 J_k^{ab}(Q_0,\modQ;T)=\sumint_q  \frac{q_0^k}{\left[ q^2-M_a^2 \right] \left[(q-Q)^2-M_b^2 \right]},
 \label{Jint}
 \end{equation}
 respectively. We work within the imaginary-time formalism of Thermal Field Theory~\cite{galekapustabook} so that, at finite temperature, momentum time-like integrals turn into Matsubara sums: 
 \begin{equation}
 \sumint_q\equiv \tsum \int \frac{d^{D-1}q}{(2\pi)^{D-1}}
 \end{equation}
 with $q_0=\omega_n\equiv 2\pi\, i\, n\, T$, $Q_0=\omega_m$ with $n,\,m$ integers, and the Euclidean metric $(-,-,-,-)$ is used.
 Note that the $J$-integrals in ${\cal T}_4^J$ come from diagrams (e), (f), and (g) in Fig.~\ref{fig:diags}, while the tadpoles in ${\cal T}_4^F$ come from diagrams (c) and (d), as well as from relations between the $J$ integrals, to be explained below. We also recall that the tadpole contributions coming from diagram (d) in Fig.~\ref{fig:diags} are encoded in the self-energy and residue of the propagator for the external legs, which are both $T$-dependent through $F_{\beta a}(T)$. That correction is only perturbatively relevant in the ${\cal T}_2$ amplitude, as explained at $T=0$, e.g., in \cite{GomezNicola:2001as}.
 
 It is important to observe that the $J$ integrals depend separately on external timelike and spacelike momenta due to the loss of Lorentz covariance in the thermal bath. Actually, while at $T=0$ energy-momentum conservation and the on-shell condition for external legs leave only three Lorentz-invariant quantities for a generic scattering process $p_a\,p_b\rightarrow p_c\,p_d$,  e.g., $p_a\cdot p_b$,  $p_a\cdot p_c$, $p_b\cdot p_c$ or $s,t,u$, at $T\neq 0$ one has six independent rotation invariants, e.g., $\vec{p_a}\cdot \vec{p_b}$, $\vec{p_a} \cdot \vec{p_c}$, $\vec{p_b}\cdot \vec{p_c}$, $p_a^0$,  $p_b^0$, $p_c^0$, which we can recast into $s,\,t,\,u,\,S^0,\,T^0,\,U^0$. In both cases, the on-shell $s+t+u=\sum M_i^2$ condition reduces the number of independent variables to two and five, respectively. 
 
 Once the Matsubara sums are performed using standard Thermal Field Theory, the resulting integrals can be analytically continued for scattering processes through   
 \begin{equation}
 J_k (2\pi mT,\modQ)\rightarrow (-i)^k J_k(-i(Q_0+i\epsilon),\modQ),
 \label{AC}
 \end{equation} 
  with $Q_0\in\IR$~\cite{GomezNicola:2002tn,galekapustabook}. Explicit expressions for the above $J_k^{ab}$ loop integrals and tadpoles $F_i$ at finite temperature are given in Appendix~\ref{app:loop}, where we also discuss  their most relevant analytical properties  for the present work. Note that due to the loss of Lorentz covariance in the thermal bath, there are three independent $J_k^{ab}$ functions for $k=0,\,1,\,2$. Nevertheless, in the CM frame, they are related through~\eqref{relJ1CM}-\eqref{relJ2CM}. In addition, it is worth noting that ${\cal T}_4^J$ contains the integrals responsible for elastic unitarity above the physical pion-kaon threshold. 
 
 Another important feature of the finite-temperature case is that the usual $T=0$ relations based on Lorentz covariance (Pasarino-Veltman like relations), which allows one to write loop integrals with momenta in the integrand numerator, such as $J_k$ above, in terms of just one integral, say $J_0$~\cite{Gasser:1983yg,Gasser:1984gg}, are no longer valid due to the loss of Lorentz covariance. Nevertheless, there are still useful relations between the  $F$ and $J_k$ loop thermal integrals, which help to simplify the expression for the amplitude. Those relations are also provided in Appendix~\ref{app:loop}.

The thermal amplitude can be decomposed into partial waves $t_{IJ}$ of definite isospin $I$ and angular momentum $J$. One has to be careful only with performing the usual crossing symmetry now in terms of $\Sr,\Tr,\Ur$ variables instead of $s,t,u$, which gives rise to the isospin projections:

\begin{eqnarray}
{\cal T}_{3/2} (\Sr,\Tr,\Ur;T)&=&{\cal T} (\Sr,\Tr,\Ur;T), \label{t32} \\
{\cal T}_{1/2} (\Sr,\Tr,\Ur;T)&=&\frac{1}{2}\left[3{\cal T} (\Ur,\Tr,\Sr;T)-{\cal T} (\Sr,\Tr,\Ur;T)\right], \label{t12} 
\end{eqnarray}
with ${\cal T} (\Sr,\Tr,\Ur;T)$ the finite-temperature $K^+\pi^+\rightarrow K^+\pi^+$ amplitude, which we collect in Appendix~\ref{app:amplitude}. For the partial-wave projection, we consider the center of momentum (CM) frame (see Appendix~\ref{app:kin}) and define~\cite{GomezNicola:2001as}
\begin{equation}
t_{IJ} (s;T)=\frac{1}{32\pi}\int_{-1}^1 dx P_J(x) {\cal T}_I \left[S_0=\sqrt{s},\vec{S}=\vec{0},T_0=0,\vert \vec{T} \vert^2=2p^2(s)(1-x),U_0=\frac{\Delta}{\sqrt{s}},\vert \vec{U} \vert^2=2p^2(s)(1+x);T\right]
\label{tIJ}
\end{equation}
with $x$ the cosine of the scattering angle in the CM frame, $P_J(x)$ the Legendre polynomial of order $J$, $p(s)=p_{CM}(s;M_K,M_\pi)$ defined in~\eqref{pcm} and $\Delta=\Delta_{K\pi}=M_K^2-M_\pi^2$ .

 \section{Analytical and cut structure of the thermal amplitude. Thermal unitarity and Landau cuts} 
\label{sec:anther}

In this section, we discuss the analytic structure in the complex-$s$ plane of the perturbative ChPT partial waves at finite temperature and obtain a generalized thermal unitarity relation, which includes the Landau cut contribution arising from the pion-kaon mass difference $M_K-M_\pi$. Such thermal unitarity relation generalizes the results in~\cite{GomezNicola:2002tn} for equal-mass scattering and supports the unitarization method explained in section~\ref{sec:unit}.
In Appendix~\ref{app:loop} (sections \ref{sec:Janstr} and \ref{sec:parcases}), we provide a detailed description of the analytic structure of the $J$-loop integrals in the $S$, $T$, and $U$ channels.

At nonzero $T$, the generic $J_k^{ab}(Q)$ thermal function given in~\eqref{Jint} can be integrated using standard contour techniques, leading to the results collected in~\eqref{generalJ}.  Its analytic structure, depicted in Fig.~\ref{fig:Jcuts}, can be obtained from its imaginary part in~\eqref{imJgen}. The modifications with respect to the $T=0$ case are twofold: first, the unitary cut contribution for $Q^2\equiv s\geq (M_a+M_b)^2$ is weighted by the combination of Bose-Einstein functions $1+n_1+n_2$, where $n_i$ is short for $n_B(E_i)=(e^{E_i/T}-1)^{-1}$ and the unity contribution stands for the $T=0$ part; second, a new cut, denoted as the Landau cut, appears for $-\vert \vec{Q} \vert ^2\leq s\leq (M_a-M_b)^2$ and is weighted by $n_1-n_2$.

The presence of these two cuts and their statistical Bose-Einstein  weights are explained by scattering processes coming from the interactions of particles in the thermal bath with the incoming and/or outgoing asymptotic states~\cite{GomezNicola:2002tn,Weldon:1983jn,GomezNicola:2002an,MontanaFaiget:2022cog}. 
Namely, for a generic  scattering process $a\,b\rightarrow c\,d$, one has to take into account the stimulated production of the $c$ and $d$ states from scattering with the thermal bath components, as well as their absorption by the thermal bath; the Bose-Einstein function $n_i$ and $1+n_i$ weight the probability for absorption and production of meson $i$, respectively. These possible interactions with the bath fall into two categories corresponding to the unitary and the Landau cuts, namely, processes with even and odd numbers of particles in the initial and final states. 

In more detail, denoting by bar letters the particles in the bath, the unitary-like processes are the stimulated production process $ \bar a \bar b \rightarrow c d$ (in other words, $\text{bath}\rightarrow \text{bath} + c\, d$) weighted by the function $(1+n_c)\times(1+n_d)$, and the absorption $c\,d \rightarrow \bar a\, \bar b$ ($\text{bath}+cd\rightarrow \text{bath}$) weighted by $-n_c\times n_d$. For negative energies $E_c$ and/or $E_d$, one could have instead the production process $0\rightarrow \bar a\,\bar b\, c\,d $  weighted by $(1+n_c)\times(1+n_d)$, minus absorption $\bar a\, \bar b\, c\, d\rightarrow 0$ weighted by $n_c \times n_d$. The net contribution in both cases   is $(1+n_c)\times(1+n_d)-n_c\times n_d=1+n_c+n_d$, which we readily identify as the $1+n_1+n_2$ factor associated to the unitary cut in~\eqref{imJgen}.  The Landau-like processes correspond to either $\bar a\,\bar b\, c\rightarrow d$ ($\text{bath}\rightarrow \text{bath} + d$) weighted by $n_c\times (1+n_d)$, minus the inverse process $d\rightarrow \bar a\,\bar b\, c$ ($\text{bath}+d\rightarrow \text{bath}$) weighted by $n_d\times (1+n_c)$, or the reactions $\bar a\,\bar b \,d\rightarrow c$ weighted by $n_d\times (1+n_c)$ minus $c\rightarrow \bar a\, \bar b\, d$ weighted by $n_c\times (1+n_d)$.  The net contribution is proportional to $n_c-n_d$ corresponding to the $n_1-n_2$ contribution in~\eqref{imJgen}. 

Note that not all these reactions will be allowed by the kinematics of a given scattering process (see below) and that, in the above thermal processes, an antiparticle state is understood as an incoming line changed into an outgoing one with respect to  $a\,b\rightarrow c\,d$ scattering or conversely.  
 
\begin{figure}[h]
 \centering
\includegraphics[width=11cm]{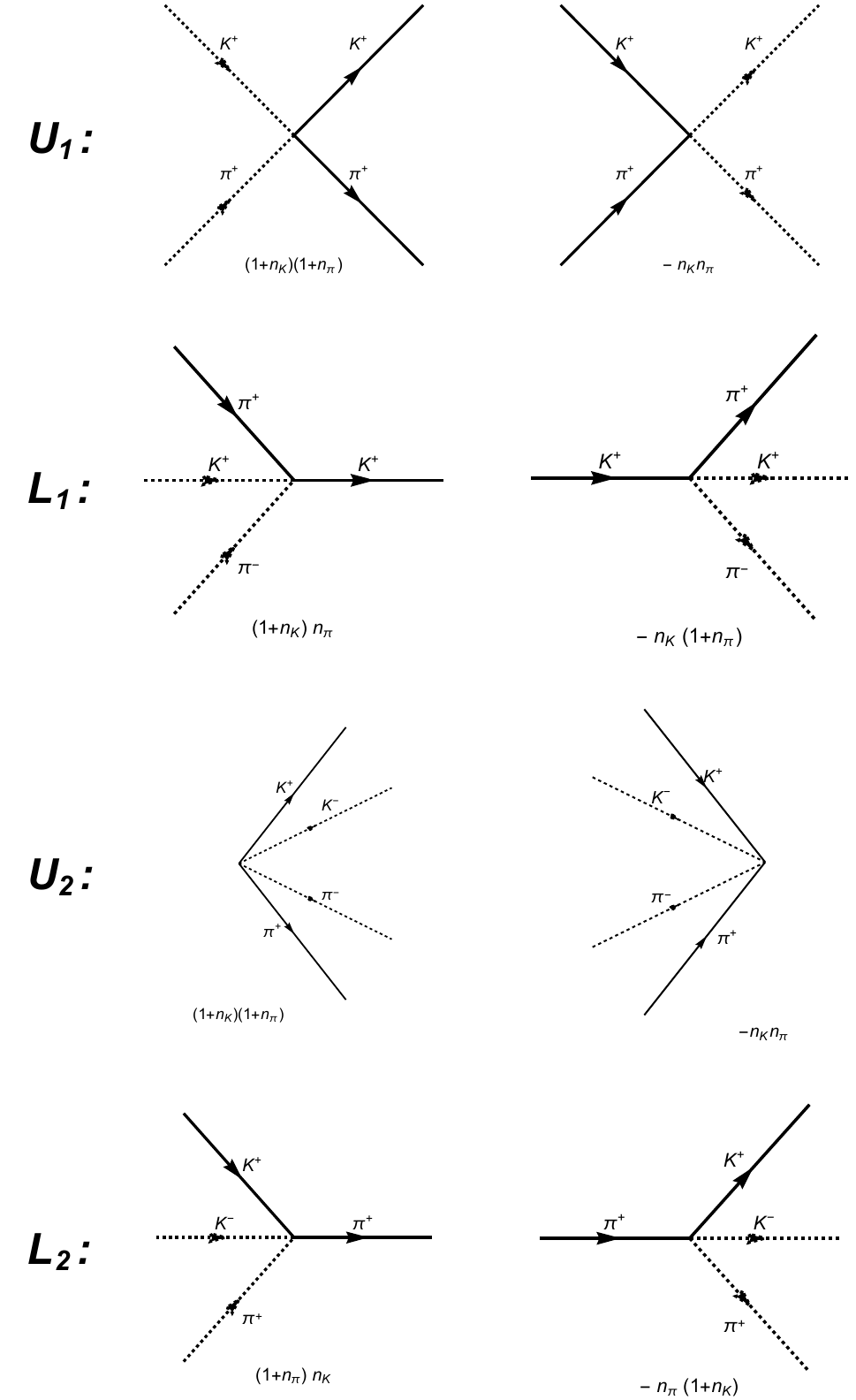}
\caption{Processes  contributing to $K^+\pi^+ \rightarrow K^+\pi^+$ scattering in the thermal bath. Dashed lines denote thermal bath particles, solid lines correspond to states produced or absorbed in the bath, while $U$ and $L$ stand for processes contributing to the Unitary or Landau cut, respectively. We also include the statistical Bose-Einstein factors associated with every process, where $n_{\pi,K}$ stand for $n_B(E_{\pi,K})$. Note that for each process of thermal production of $\pi^+,\,K^+$ outgoing states, there is an absorption process carrying a relative minus sign with respect to the production one.} 
\label{fig:proc}
\end{figure}

 As the case of interest here, the possible thermal bath processes for  $K\pi\rightarrow K\pi$ scattering  are depicted in Fig~\ref{fig:proc}. As we will immediately explain, the only allowed processes for which all of the particles involved are physical, i.e., with positive energies, are those labeled as $U_1$ and $L_1$ in the figure.

\begin{figure}[h]
 \centering
\includegraphics[width=12cm]{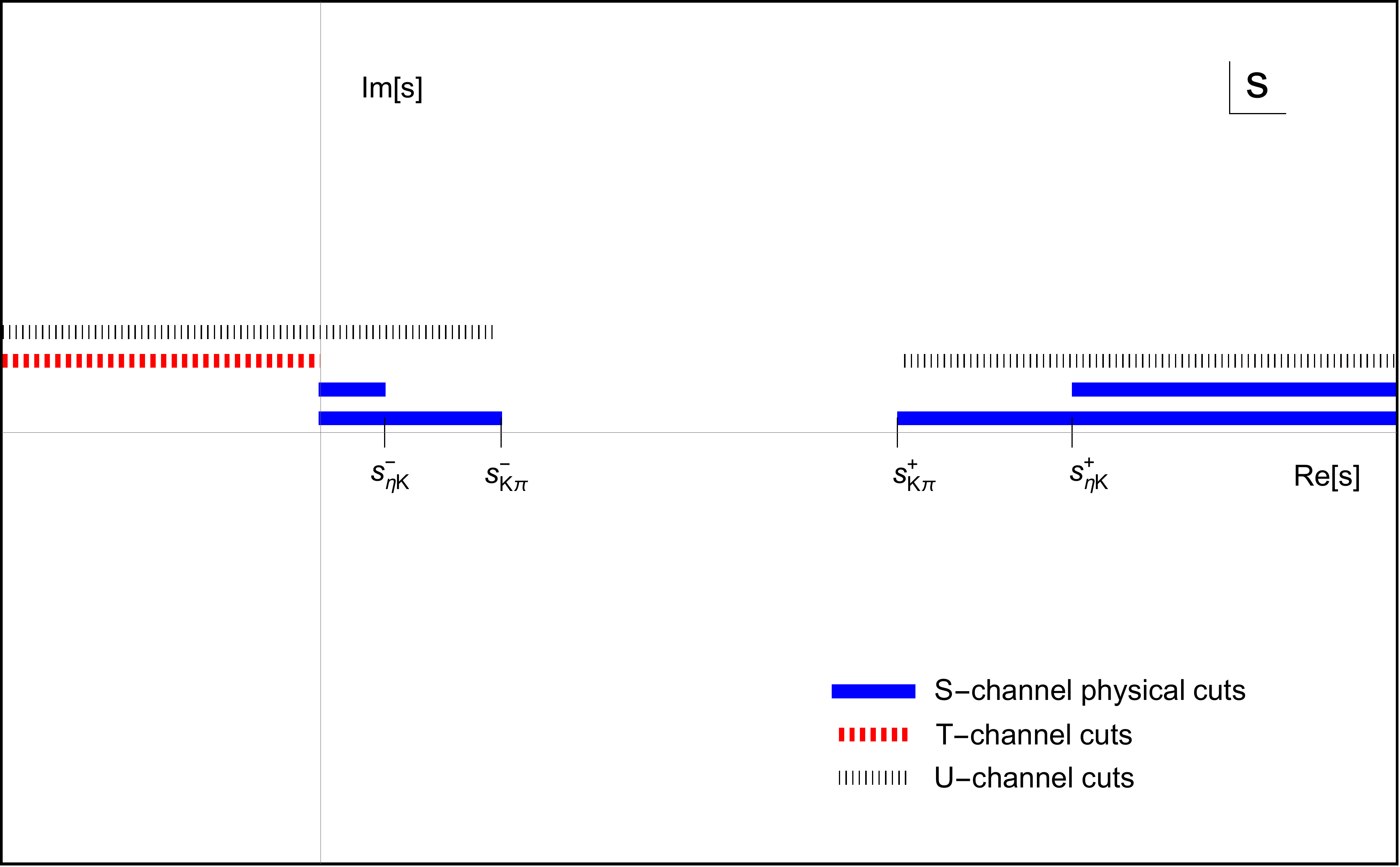}
\caption{General cut structure of $\pi K$ scattering partial waves at finite temperature with $s_{ab}^{\pm}=(M_a\pm M_b)^2$. Note that in the $I=3/2$ channel, the physical cuts opening above $s>s^+_{\eta k}$ and $s<s^-_{\eta k}$ are not present. The circular cut discussed in section \ref{sec:circularcut} is not displayed here.}
\label{fig:ampcuts}
\end{figure}

In the following, we will provide the explicit expressions for the imaginary part of the $\pi K$ partial waves alongside the $S$-channel physical cuts including thermal scattering processes and the position of the unphysical $T,\,U$-channel cuts.  We have summarized in Figure~\ref{fig:ampcuts} the different discontinuities contributing to the $\pi K$ scattering partial waves arising from this analysis. 

\subsection{Physical $S$-channel cuts: thermal unitarity}
\label{sec:Scuts}

For a given isospin channel,  the $J_k(S)$ thermal integrals provide the $S$-channel discontinuity or physical cut, which is related to the physical processes taking place in the thermal bath and renders the thermal unitarity relation, as we are about to see. 

At $T=0$, one has only the standard unitary cuts starting from each two-particle scattering threshold. For $\pi K$ scattering, the first such cut opens at $s\geq (M_K+M_\pi)^2$ coming from the $J^{K\pi} (S)$ integrals. For the $I=1/2$ channel, a second physical cut (inelastic) starts at $s\geq (M_K+M_\eta)^2$, corresponding to the process $K\pi\rightarrow K\eta$ and coming from the $J^{\eta K} (S)$ loop function. Note that due to this inelastic contribution, the unitarization of the $I=1/2$ channel would require a coupled-channel approach ~\cite{GomezNicola:2001as,Ledwig:2014cla}. However, the main properties of the $K_0^*(700)$ and $K^*(890)$ resonances can be understood by considering only the elastic region~\cite{Pelaez:2021dak,Pelaez:2016klv} and we will follow the same approach here. 

At nonzero $T$, we have, on the one hand, the thermal correction to $\im J^{K\pi}(S)$ in the unitary cut. For the $K^+\pi^+ \rightarrow K^+\pi^+$ process, i.e., the $I=3/2$ one, the thermal bath processes contributing to such modification are those labeled as $U_{1,2}$ in~Figure~\ref{fig:proc}. However, it is easy to see that the $U_2$-like processes require negative energies for at least one of the incoming (outgoing) states since the sum of all particle energies has to vanish. Therefore, the only physical processes contributing to the unitary cut at finite temperature are the $U_1$ ones. Note that in the CM frame, these processes give rise to the first contribution to the imaginary part of $J^{K\pi}(S)$ in~\eqref{imJs}, with $Q_0=E_K+E_\pi=\sqrt{s}>0$, $E_K=\frac{s+\Delta_{K\pi}}{\sqrt{s}}$, $E_\pi=\frac{s-\Delta_{K\pi}}{\sqrt{s}}$, which are both positive since $s>\Delta_{K\pi}>0$ at the unitary cut. 

 On the other hand, the Landau cut is generated by the  processes $L_{1,2}$ in Figure~\ref{fig:proc}. For $L_1$, we have $\bar E_K + \bar E_\pi= E_K-E_\pi=\sqrt{s}$, where the bars correspond to the thermal bath particles; i.e, we are denoting  $s=(p_K-p_\pi)^2=(\bar p_K + \bar p_\pi)^2$ so that in the CM frame, $\vec{p_\pi}-\vec{p_K}=\vec{0}$.  With positive energies of the thermal bath particles, the solution to the energy conservation equation in the CM frame is $E_K=\frac{s+\Delta_{K\pi}}{\sqrt{s}}$ and $E_\pi=\frac{\Delta_{K\pi}-s}{\sqrt{s}}$ with $Q_0=E_K+E_\pi$, which satisfy $E_K>E_\pi>0$ since $0< s < (M_K-M_\pi)^2<\Delta_{K\pi}$ for the Landau cut. These processes provide the second contribution to $\im J^{K\pi}(S)$~in \eqref{imJs}. For the $L_2$-like processes, we have $\bar E_K + \bar E_\pi= E_\pi-E_K$, which cannot be satisfied for all positive energies since $M_\pi<M_K$ and they do not contribute to the imaginary part of $J^{K\pi}(S)$.

The crucial point for obtaining the thermal unitarity relation at one-loop order in ChPT is to realize that at $T=0$,
\begin{equation}
\im t^{IJ}(s)=\im t^{IJ}_4(s)=16\pi\,\im J_0^{K\pi} (S)\,\left[t_2^{IJ}(s)\right]^2=\sigma_{K\pi}(s)\left[t_2^{IJ}(s)\right]^2,
\end{equation}
where $t_2^{IJ}$ and $t_4^{IJ}$ stand for the $\Od(p^2)$ and $\Od(p^4)$ ChPT amplitudes, respectively, and $\sigma_{K\pi}(s)=(2/\sqrt{s})\,p_{\text CM}(s;M_k,M_\pi)$ is the two-body phase space. It implies that, for any $IJ$ channel, $\im J_0^{K\pi} (S)$ in the $t_4$ ChPT amplitude comes multiplied by $16\pi\left[t_2^{IJ}(s)\right]^2$.  
At finite temperature, once the relations \eqref{relJ1CM} and \eqref{relJ2CM} are used, the $T$-independent function multiplying $J_0(s;T)$ is the $T=0$ one. In addition, we have to take into account those processes allowed in the thermal bath. For instance, the process $K^+\pi^+\pi^-\rightarrow K^+$ with momenta $p_K p_\pi \bar p'_\pi\rightarrow p'_K$, which can be obtained from the $K^+\pi^+\rightarrow K^+\pi^+$ reaction by changing $\bar p'_\pi\rightarrow -p'_\pi$. 
This transformation leaves the $t_2$  amplitude  invariant in terms of the variables $s=(p_K+p_\pi)^2=(p'_K-p'_\pi)^2$, $t=(p_K-p'_K)^2=(p_\pi+p'_\pi)^2$, and $u=(p_K+p'_\pi)^2=(p'_K-p_\pi)^2$. The same argument applies to the thermal amplitude in terms of $s,\,t,\,u,\,S^0,\,T^0,$ and $U^0$ variables. Thus, an extended thermal perturbative unitarity relation can be obtained by including the physical Landau-type process multiplied by its corresponding thermal phase-space factor~\eqref{imJs} for that cut.  In the $I=1/2$ channel, one also has to take into account the contribution from the $\eta K$ intermediate sate, which involves $\im J_0^{\eta K} (S)$, this time multiplied by $\left[t_2^{K\pi\rightarrow K\eta}\right]^2$. Nevertheless, one arrives to the same conclusion regarding the inclusion of the thermal bath processes $\bar K_0\pi^0\rightarrow \bar K_0\eta$ and $\bar K_0 K_0\pi^0\rightarrow \eta$. 

Altogether, the generalized thermal unitarity relation for the perturbative $K\pi\rightarrow K\pi$ ChPT amplitude for the $I=3/2,\,J=0,1$ channels reads:

\begin{eqnarray}
\im t_4^{K\pi\rightarrow K\pi}(s;T)&=&\sigma_{K\pi}^T(s) \left[t_2^{K\pi\rightarrow K\pi}(s)\right]^2 \theta\left[s-(M_K+M_\pi)^2\right] + 
\tilde\sigma_{K\pi}^T(s;T) \left[t_2^{K\pi\pi\rightarrow K}(s)\right]^2 \theta\left[(M_K-M_\pi)^2-s\right]\theta\left(s\right)
\nonumber\\
&=& \left[t_2^{K\pi\rightarrow K\pi}(s)\right]^2 \left\{ 
\sigma_{K\pi}^T(s) \,\theta\left[s-(M_K+M_\pi)^2\right] + 
\tilde\sigma_{K\pi}^T(s;T)\,  \theta\left[(M_K-M_\pi)^2-s\right] \theta\left(s\right)
\right\},
\label{thermalpertunit32}
\end{eqnarray}
while for $I=1/2,J=0,1$ one has

\begin{eqnarray}
\im t_4^{K\pi\rightarrow K\pi}(s;T)&=&\sigma_{K\pi}^T(s) \left[t_2^{K\pi\rightarrow K\pi}(s)\right]^2 \theta\left[s-(M_K+M_\pi)\right]^2
+\sigma_{\eta K}^T(s) \left[t_2^{K\eta\rightarrow K\pi}(s)\right]^2 \theta\left[s-(M_K+M_\eta)\right]^2
\nonumber \\
&+& 
\tilde\sigma_{K\pi}^T(s) \left[t_2^{K\pi\pi\rightarrow K}(s)\right]^2 \theta\left[(M_K-M_\pi)^2-s\right]\theta\left(s\right)+ 
\tilde\sigma_{\eta K}^T(s) \left[t_2^{KK\pi\rightarrow \eta}(s)\right]^2 \theta\left[(M_\eta-M_K)^2-s\right]\theta\left(s\right)
\nonumber\\
&=&
\left[t_2^{K\pi\rightarrow K\pi}(s)\right]^2\left\{\sigma_{K\pi}^T(s)\,  \theta\left[s-(M_K+M_\pi)\right]^2+
\tilde\sigma_{K\pi}^T(s)\,  \theta\left[(M_K-M_\pi)^2-s\right]\theta\left(s\right)
\right\}
\nonumber\\
&+& \left[t_2^{K\eta\rightarrow K\pi}(s)\right]^2 \left\{  \sigma_{\eta K}^T(s)\, \theta\left[s-(M_K+M_\eta)\right]^2
+ \tilde\sigma_{\eta K}^T(s)\, \theta\left[(M_\eta-M_K)^2-s\right]\,\theta\left(s\right)
\right\},
\label{thermalpertunit12}
\end{eqnarray} 
with the thermal phase space factors
\begin{eqnarray}
\sigma_{ab}^T(s)&=&\sigma_{ab}(s)\left[ 1+n_B\left(\frac{s+\Delta_{ab}}{2\sqrt{s}}\right)+n_B\left(\frac{s-\Delta_{ab}}{2\sqrt{s}}\right)\right],
\label{therps}\\
\tilde\sigma_{ab}^T(s)&=&\sigma_{ab}(s)\left[ n_B\left(\frac{\Delta_{ab}-s}{2\sqrt{s}}\right)-n_B\left(\frac{s+\Delta_{ab}}{2\sqrt{s}}\right)\right],
\label{therpstilde}
\end{eqnarray}
with $\Delta_{ab}=M_a^2-M_b^2$, so that the above equations  can be used to easily generalize thermal unitarity for any thermal scattering amplitude.

\subsection{$T$-channel cuts}
\label{sec:Tcuts}

The analytic properties of the $J(0,\sqrt{-t(s,\theta)};T)$ thermal integrals with $t(s,\theta)$ in \eqref{TCM}, are collected in Appendix~\ref{app:t-channel}, where one can see that, in the CM frame, the only imaginary part arises for $s<0$, and the range of the scattering angle where  $(M_a+M_b)^2<t(s,\theta)<-4\,p_{CM}^2(s)$. We remind that for the $\pi K$ scattering amplitude, only $J^{ab}(T)$ integrals with $M_a=M_b$ arise.

\subsection{$U$-channel cuts}
\label{sec:Ucuts}

The $J(U)$ integrals contributing to the $\pi K$ partial waves~\eqref{t32}~and~\eqref{t12} come from the loop-functions $J_k^{K\pi}(U)$ and $J_k^{\eta K}(U)$. In the CM frame, taking into account \eqref{UCM}, we can see that the Mandelstam variable $u(s,\theta)$ is bounded by 
\begin{align}
-s+2\Sigma_{K\pi}\leq u \leq \Delta_{K\pi}^2/s,\quad\text{for}\quad p^2_{\text CM}(s;\,M_K,M_\pi)\geq 0,\\\nonumber\\
 \Delta_{K\pi}^2/s \leq u \leq -s+2\Sigma_{K\pi},\quad\text{for}\quad   p^2_{\text CM}(s;\,M_K,M_\pi)\leq 0,\nonumber
\end{align}
since the cosine of the scattering angle varies between $-1$ and 1, and where we have used  $$\Delta_{K\pi}^2/s-4p^2_{\text CM}(s;M_K,M_\pi)=-s+2\Sigma_{K\pi},\quad \Sigma_{K\pi}=M_K^2+M_\pi^2 .$$ 

First, for $s\geq (M_K+M_\pi)^2$, and taking into account that in that region
 \begin{equation}
 p^2_{\text{CM}}(s;M_K,M_\pi)\geq 0\quad\text{and}\quad
 \Delta_{K\pi}^2/s\leq (M_K-M_\pi)^2\leq (M_K+M_\pi)^2\leq (M_K+M_\eta)^2,    
 \end{equation}
it can be seen that the $J^{K\pi}(U)$ and $J^{\eta K}(U)$ loop functions develop a Landau cut but not the unitary one. Note that in this region $u \geq -\vert \vec{U} \vert^2$ since $U_0^2=\Delta_{K\pi}^2/s>0$ and that $u$ can take any value in the range $-\vert \vec{U} \vert^2<u<0$ for arbitrarily large values of $s$.  
One of the most important conclusions of this result is that this unphysical $U$-channel discontinuity overlaps with the physical unitary cuts corresponding to the $K\pi\rightarrow K\pi$ and $K\pi\rightarrow K\eta$ processes discussed in Section~\ref{sec:Scuts}, as shown in Fig.~\ref{fig:ampcuts}, which is a genuine thermal effect. 

Second, in the interval $(M_K-M_\pi)^2\leq s \leq (M_K+M_\pi)^2$, where $p^2_{\text CM}(s;\,M_K,M_\pi)\leq 0$, there is no Landau-cut contribution either from $J^{K\pi}(U)$ or from $J^{\eta K}(U)$ since $\Delta_{K\pi}^2/s\geq (M_K-M_\pi)^2 \geq (M_\eta-M_K)^2$. In addition, the maximum value of $-s+2\Sigma_{K\pi}$ in this interval is $(M_K+M_\pi)^2<(M_\eta+M_K)^2$ reached at $s=(M_K-M_\pi)^2$; hence, these values of $u$ lie also outside the unitary cut. It implies that the amplitude remains real in that interval, which allows one to apply Schwarz's reflection principle for the analytic continuation of the different partial waves.

 Third, in the interval $0\leq s \leq (M_K-M_\pi)^2$, $p^2_{\text CM}(s;\,M_K,M_\pi)\geq 0$ and 
$\Delta_{K\pi}^2/s\geq (M_K+M_\pi)^2$, so that $u$ takes arbitrarily large values near $s=0$. It implies that the unitary cuts of both $J^{K\pi}(U)$ and $J^{\eta K}(U)$ are reached for $\cos\theta$ sufficiently close to $-1$. On the contrary,  the minimum value of $-s+2\Sigma_{K\pi}$ is $(M_K+M_\pi)^2>(M_K-M_\pi)^2>(M_\eta-M_K)^2$, so that  $J^{K\pi}(U)$ and $J^{\eta K}(U)$ do not develop Landau cuts. 

Finally, for $s<0$, $p^2_{\text CM}(s;\,M_K,M_\pi)\leq 0$ and $-s+2\Sigma_{K\pi}>2\Sigma_{K\pi}>(M_K+M_\pi)^2$ so that both 
 unitary cuts of $J^{K\pi}(U)$ and $J^{\eta K}(U)$ are reached, while $U_0^2=\Delta_{K\pi}^2/s<0$ so that $u$ lies off the Landau cuts for those integrals.

\subsection{Circular cut}
\label{sec:circularcut}

So far, we have discussed the discontinuities of the  one-loop $\pi K$ amplitude ${\cal T} (\Sr,\Tr,\Ur;T)$. Nevertheless, when dealing with pion-kaon partial waves, defined in~\eqref{tIJ}, there is an additional source of singularities one has to take into account, which involves the angular integration over the Legendre polynomials $P_J$ in \eqref{tIJ}. 
The easiest way to analyze these singularities is through the so-called Froissart-Gribov representation of scattering partial waves, which, in turn, is expressed in terms of the Legendre functions of second kind $Q_J(x)$. The functions $Q_J(x)$ are analytic in the $x$ plane but for two discontinuities with branch points at $x=\pm 1$. Thus, in the CM frame, pion-kaon partial waves will be singular at those values of $s$ satisfying
\begin{align}
1+\frac{t}{2 p_{\text CM}^2}=\pm 1,\quad 1+\frac{1}{2 p_{\text CM}^2}\left(u-\frac{\Delta_{\pi K}^2 }{s}\right)=\pm 1,\label{circ_cut_1}
\end{align}
for values of $t$ and $u$ in the ranges
\begin{align}
4M_\pi^2 \le t, \quad (M_\pi+M_K)^2 \le u,\label{circ_cut_2}
\end{align}
and  $p_{\text CM}$ the momentum in the CM frame defined in~\eqref{pcm}.
The solutions of~\eqref{circ_cut_1} and~\eqref{circ_cut_2} involve two additional discontinuities. Namely, around the circle $\vert s \vert \le M_K^2-M_\pi^2$ and along the negative real axis from $s \in (-\infty,\,0]$, arising from the $t$ channel; along the real axis from $(M_K-M_\pi)^2$ to $-\infty$, arising from the $u$ channel. 
Nevertheless, note that none of these singularities are associated with the thermal bath; hence, they are also present at $T=0$. 

\section{Results for ChPT partial waves and scattering lengths}
\label{sec:numchpt}

In this section, we provide the numerical results from the perturbative ChPT calculation of the $\pi K$ scattering amplitude at finite temperature. 

For our calculations we have used: $M_\pi=139.6$ MeV, $M_K=495.8$ MeV, $M_\eta=552.5$ MeV, $F_\pi=92.3$ MeV, $F_K=110.6$ MeV, $F_{\eta}=119.2$ MeV and the LECs extracted from~\cite{Molina:2020qpw} given in Table~\ref{LECsfit}. 
In addition, all error bands shown below are obtained through the propagation in quadrature of the LECs uncertainties. 

\begin{table}[h!]
\renewcommand{\arraystretch}{1.6}
\centering
\begin{tabular}{lllllllll}
\cline{2-9}
\multicolumn{1}{c|}{}  & \multicolumn{1}{c|}{$L_1^r$} & \multicolumn{1}{c|}{$L_2^r$} & \multicolumn{1}{c|}{$L_3^r$} & \multicolumn{1}{c|}{$L_4^r$}& \multicolumn{1}{c|}{$L_5^r$}& \multicolumn{1}{c|}{$L_6^r$}& \multicolumn{1}{c|}{$L_7^r$} & \multicolumn{1}{c|}{$L_8^r$}\\ \hline
\multicolumn{1}{|c|}{LECs $\times 10^{3}$} & \multicolumn{1}{c|}{$0.70^{+0.01}_{-0.01}$} & \multicolumn{1}{c|}{$1.04$} & \multicolumn{1}{c|}{$-3.44^{+0.04}_{-0.04}$} & \multicolumn{1}{c|}{$-0.08^{+0.03}_{-0.04}$} &\multicolumn{1}{c|}{$0.98^{+0.07}_{-0.05}$} & \multicolumn{1}{c|}{$0.24^{+0.08}_{-0.06}$} & \multicolumn{1}{c|}{$0.008^{+0.090}_{-0.140}$} & \multicolumn{1}{c|}{$0.098^{+0.100}_{-0.110}$} \\ \hline                  
\end{tabular}
\caption{$\Od(p^4)$ chiral parameters ($\times10^3$) obtained in the Global Fit IV in~\cite{Molina:2020qpw}.}
\label{LECsfit}
\end{table}

\begin{figure}[h!]
 \centering
\includegraphics[width=8cm]{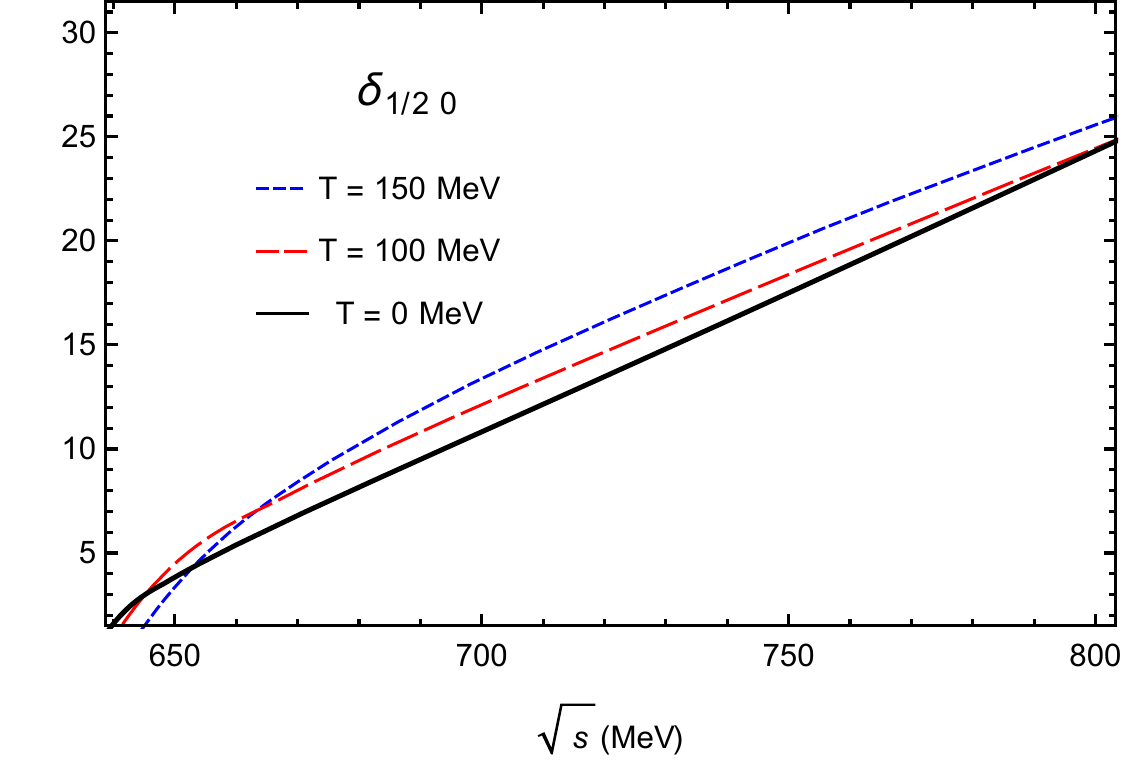}
\includegraphics[width=8cm]{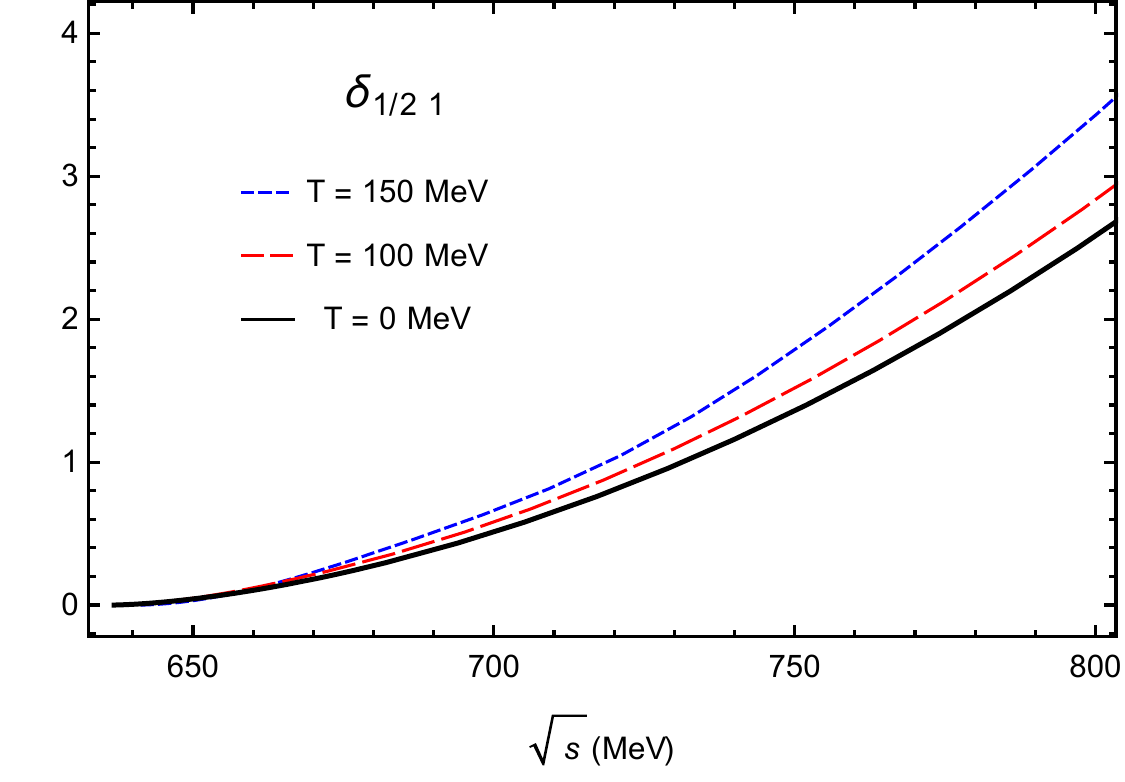}\\
\includegraphics[width=8cm]{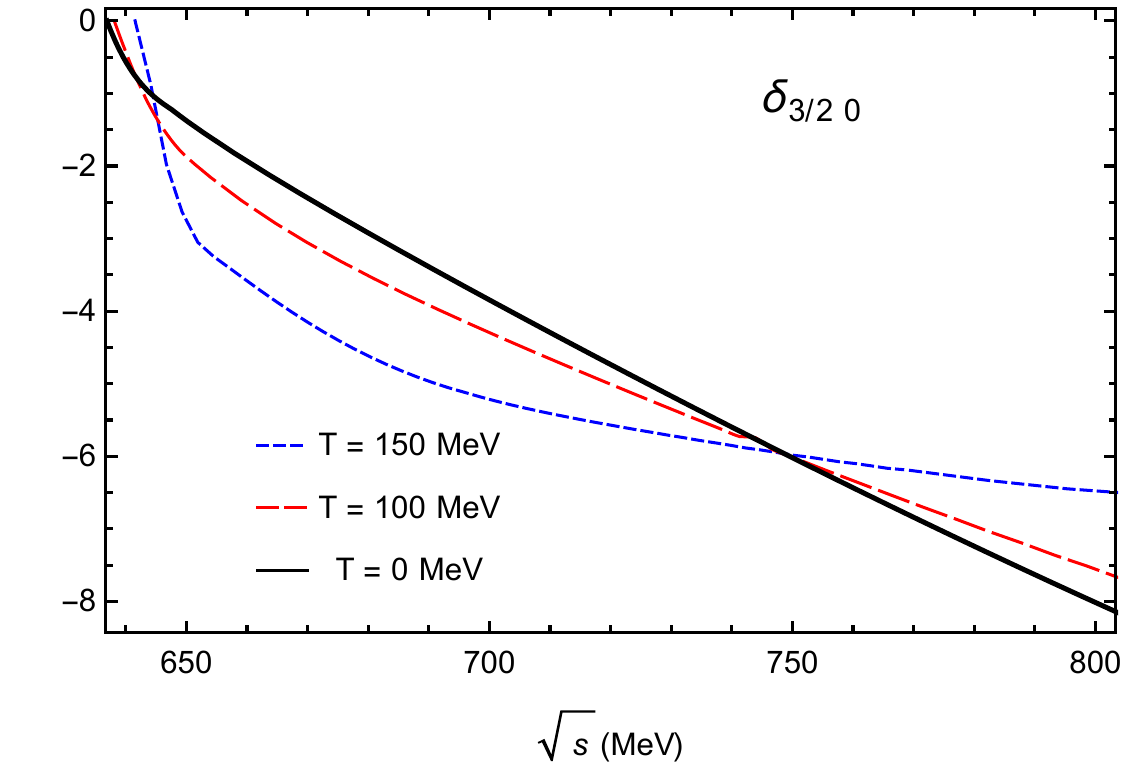}
\caption{Temperature evolution of the phase shifts $\delta_{IJ}$ for $IJ=$ $1/2$ $0$ (left), $IJ=$ $1/2$ $1$ (right) and $IJ=$ $3/2$ $0$ (bottom).}
\label{fig:des}
\end{figure}

From the ChPT $\pi K$ thermal amplitude, we can obtain the thermal evolution of any low-energy observables. For instance, we show in Fig.~\ref{fig:des} the results for the $\pi K$ phase shifts of the $S$ and $P$ waves, which at low energies can be defined as $\delta_{IJ}=\sigma_{\pi K}^T\text{Re}(t_{IJ})$.\footnote{Note that in the elastic region a partial wave is parameterized as $t_{IJ}(s)=\sin\delta_{IJ}(s){\text e}^{i\delta_{IJ}(s)}/\sigma_{IJ}(s)$, so that, at low energies, one has, $\sigma_{IJ}(s)\,\text{Re}\,t_{IJ}(s)=\sin2\delta_{IJ}(s)/2\sim \delta_{IJ}(s).$} 

The thermal evolution of the $I=1/2$, $J=0$, and $J=1$ phase shifts follow a similar trend as the $I=J=0$ and $I=J=1$ waves in $\pi\pi$ scattering, respectively, obtained in~\cite{GomezNicola:2002tn}. Namely, the phase shifts increase with temperature at low energies, reflecting the increase in the interaction strength due to the thermal bath. On the contrary, the $I=3/2,J=0$ phase shift becomes more repulsive (negative) at higher temperatures, at least at low energies where the ChPT expansion is well-behaved, hence resembling the thermal tendency of the $IJ=20$ wave in $\pi\pi$ scattering.  Note that we have considered the thermal dependence of the masses in the initial and final asymptotic states. As a consequence, there is a shift of the pion-kaon threshold at $T\neq0$ in Fig. \ref{fig:des}.

\begin{figure}[h!]
 \centering
\includegraphics[width=8cm]{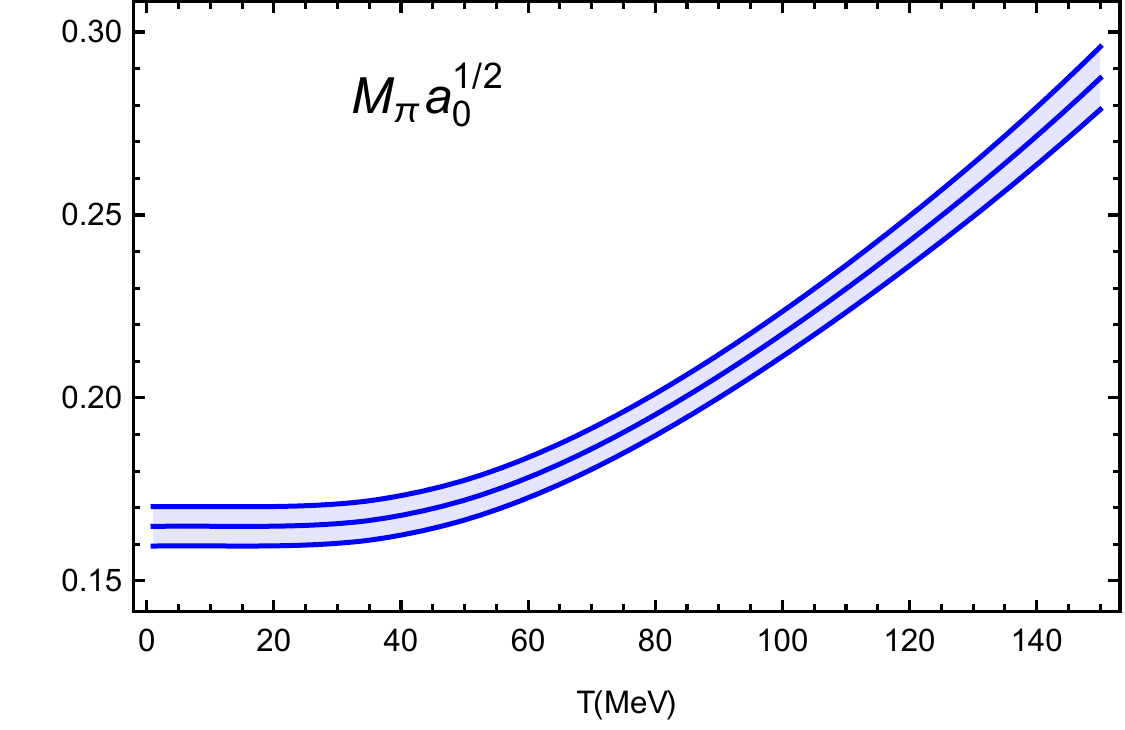}
\includegraphics[width=8cm]{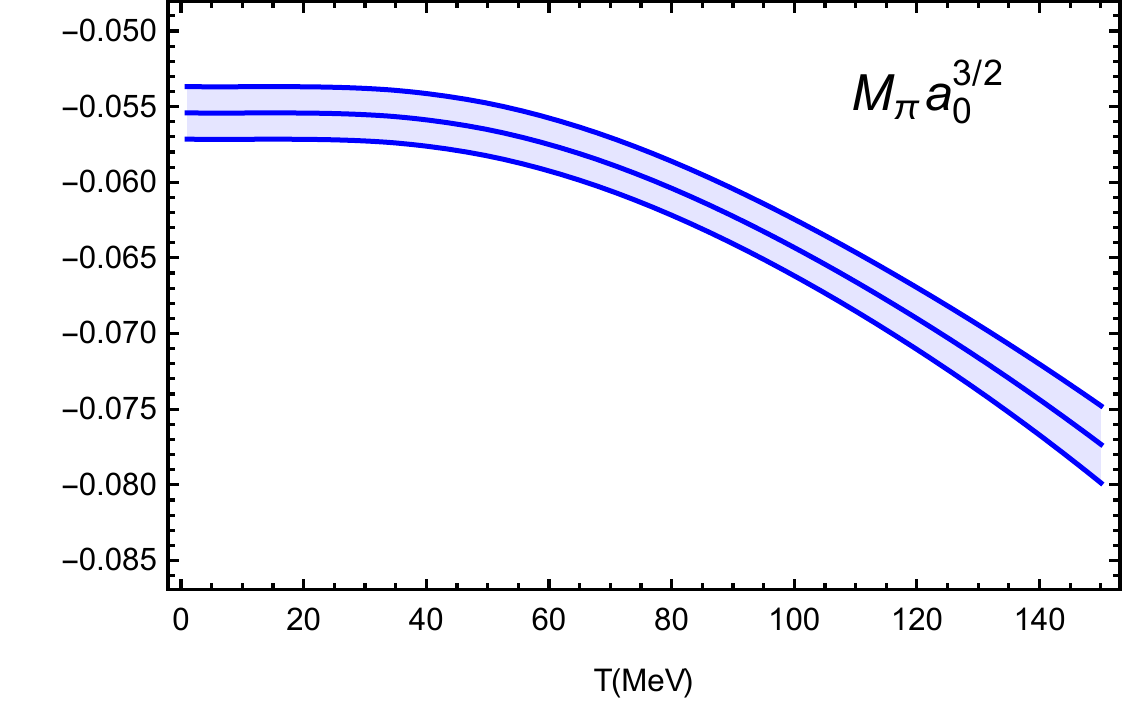}
\caption{Temperature evolution of the scattering lenghts $a^{I}_{J}$ for $IJ=$ $1/2$ $0$ (left) and $IJ=$ $3/2$ $0$ (right). The bands correspond to the LECs uncertainties, as explained in the text.}
\label{fig:scattlen}
\end{figure}

We also provide in Fig.~\ref{fig:scattlen} the temperature dependence of the corresponding scattering lengths, defined as~\cite{Dobado:1996ps,Buettiker:2003pp}
$$a_0^{I}=\frac{2}{M_\pi+M_K}\text{Re}\,t_{I0}((M_\pi+M_K)^2,T).$$

Their thermal evolution follows similar trends to those discussed above for the partial waves, highlighting that the strength of the interaction increases in the thermal bath as the temperature rises, for low and moderate energies where the ChPT approach is reliable.

\section{Unitarization and the thermal $K_0^* (700)$ and $K^*(892)$ poles}
\label{sec:unit} 

So far we have been dealing with the perturbative ChPT $\pi K$ thermal amplitude, satisfying the generalized thermal unitarity relation for partial waves derived in~\eqref{thermalpertunit32}~and~\eqref{thermalpertunit12}.  Our purpose in this section is to construct a unitarized thermal amplitude $t_U$ through the so-called Inverse Amplitude Method (IAM), developed at $T=0$ in \cite{Dobado:1996ps,GomezNicola:2001as} and for $T\neq 0$ and $\pi\pi$ scattering in~\cite{Dobado:2002xf}. 

At $T=0$, the IAM unitarized amplitude $t_U$ is constructed by demanding exact unitarity in the physical region, i.e., for any partial wave and in the elastic approximation one should have 
\begin{equation}
\im t_U(s) = \sigma(s) \vert t_{U}(s) \vert ^2 \Rightarrow \im\left(\frac{1}{t_{U}(s)}\right)=-\sigma(s), \quad s\geq (M_a+M_b)^2,
\label{invunit}
\end{equation}
with $\sigma$ the phase space, i.e., in the elastic approximation, the imaginary part of the inverse amplitude $1/t$ is completely fixed by unitarity.  The IAM amplitude is  constructed by demanding the previous condition on $\im (1/t)$ exactly and imposing that the low-energy expansion of the $\re(1/t)$ of the unitarized amplitude reproduces the ChPT prediction. That leads to $t_U(s)=\left[t_2(s)\right]^2/\left[t_2(s)-t_4(s)\right]$.

As in previous works~\cite{Dobado:2002xf}, extending now the unitarized amplitude to finite temperature with the replacement $t_4(s)\rightarrow t_4(s;T)$ with the finite-$T$ amplitude calculated in previous sections, we obtain

\begin{equation}
t_U (s;T)=\frac{\left[t_2 (s)\right]^2}{t_2(s)-t_4(s;T)},
\label{IAM}
\end{equation}
where, as explained before, we are sticking to the one-channel case for the $\pi K$ amplitude, i.e., we are demanding exact unitary only below the $K\eta$ threshold, the relevant kinematic region to reproduce the behavior of the $K_0^* (700)$ and $K^*(892)$ resonances. 

Note that the unitarized partial wave in~\eqref{IAM} satisfies $\im(1/t_U)=-\im t_4/t_2^2$, so that, from~\eqref{thermalpertunit32}~and~\eqref{thermalpertunit12}, we see that the unitarized amplitude satisfied exact (single channel) thermal unitarity, as expected from our generic definition of a thermal amplitude, i.e,

\begin{equation}
\im t_U(s;T)=\left\{
\begin{array}{l}
\sigma_{ab}^T(s) \left[t_U(s;T)\right]^2,  \quad s\geq (M_a+M_b)^2, \\ \\
\tilde\sigma_{ab}^T(s) \left[t_U(s;T)\right]^2, \quad 0\leq s \leq (M_a-M_b)^2,
\end{array}
\right.
\label{exactunit}
\end{equation}
with the thermal phase-space factors given in \eqref{therps} and \eqref{therpstilde}. The relations ~\eqref{exactunit} are then inherited from the thermal ChPT perturbative relations,  including the new Landau-type contributions coming from scattering in the thermal bath. Furthermore, $t_U=t_2+t_4+\dots$ perturbatively, as requested. 

It is also important to stress that except for the possible presence of poles in $t^{-1}(s;T)$ corresponding to zeros of $t(s,T)$, the cut structure of the unitarized amplitude is the same as that of the perturbative one depicted in Fig.~\ref{fig:ampcuts}, 
which on the $S$-like physical cuts turns into the thermal unitarity relation just discussed. However, although the $T$ and $U$ cuts remain in the same place, the imaginary part of $t_U$  and $t_4$ only coincide in the left-hand cut perturbatively, as it also happens at $T=0$~\cite{GomezNicola:2001as}.  Finally, let us recall that the second Riemann sheet, where resonance poles appear, can be defined from the first one as
\begin{equation}
t_U^{II}(s;T)=\frac{t_U(s;T)}{1+2\,i\,\sigma^{T}_{K\pi}(s)t_U(s;T) },
\end{equation}
where the $\sigma^{T}_{K\pi}(s)$ determination is chosen so that $\sigma^{T}_{K\pi}(s^*) =-\left(\sigma^{T}_{K\pi}\right)^*$ to ensure the Schwarz reflection symmetry in the second Riemann sheet. 
Having the correct analytic structure, the unitarized $\pi K$ partial wave in the second Riemann sheet can be continued into the complex plane looking for resonance poles, which at $T=0$ allows one to reproduce the $K_0^* (700)$ and $K^* (892)$ states in the $I=1/2, J=0$ and $I=1/2,J=1$ channels, respectively~\cite{Pelaez:2016klv,Pelaez:2021dak,Pelaez:2003xd,Pelaez:2020uiw,Pelaez:2020gnd}. Likewise, we can now obtain the temperature dependence of the pole position of those resonances in the second Riemann sheet, parameterized as customary as $s_{p}(T)=\left[M_{p}(T)-i\Gamma_{p} (T)/2\right]^2$, where $M_p$ and $\Gamma_p$ would approximately correspond to the mass and width of a resonance in the Breit-Wigner limit $\Gamma_p\ll M_p$, which in the present analysis would apply only to the $K^* (892)$. 
\begin{figure}[h]
 \centering
\includegraphics[width=8cm]{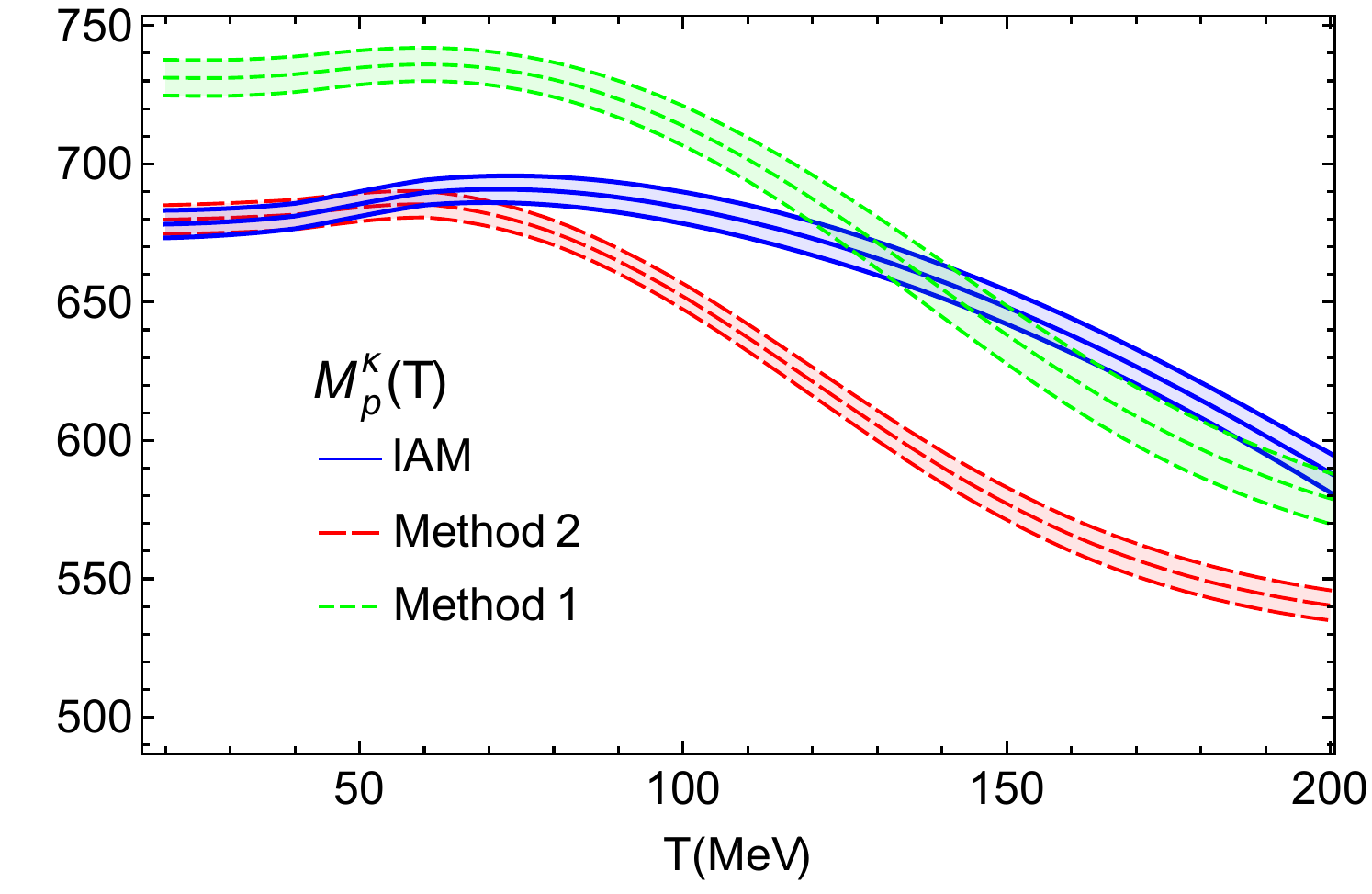}
\includegraphics[width=8cm]{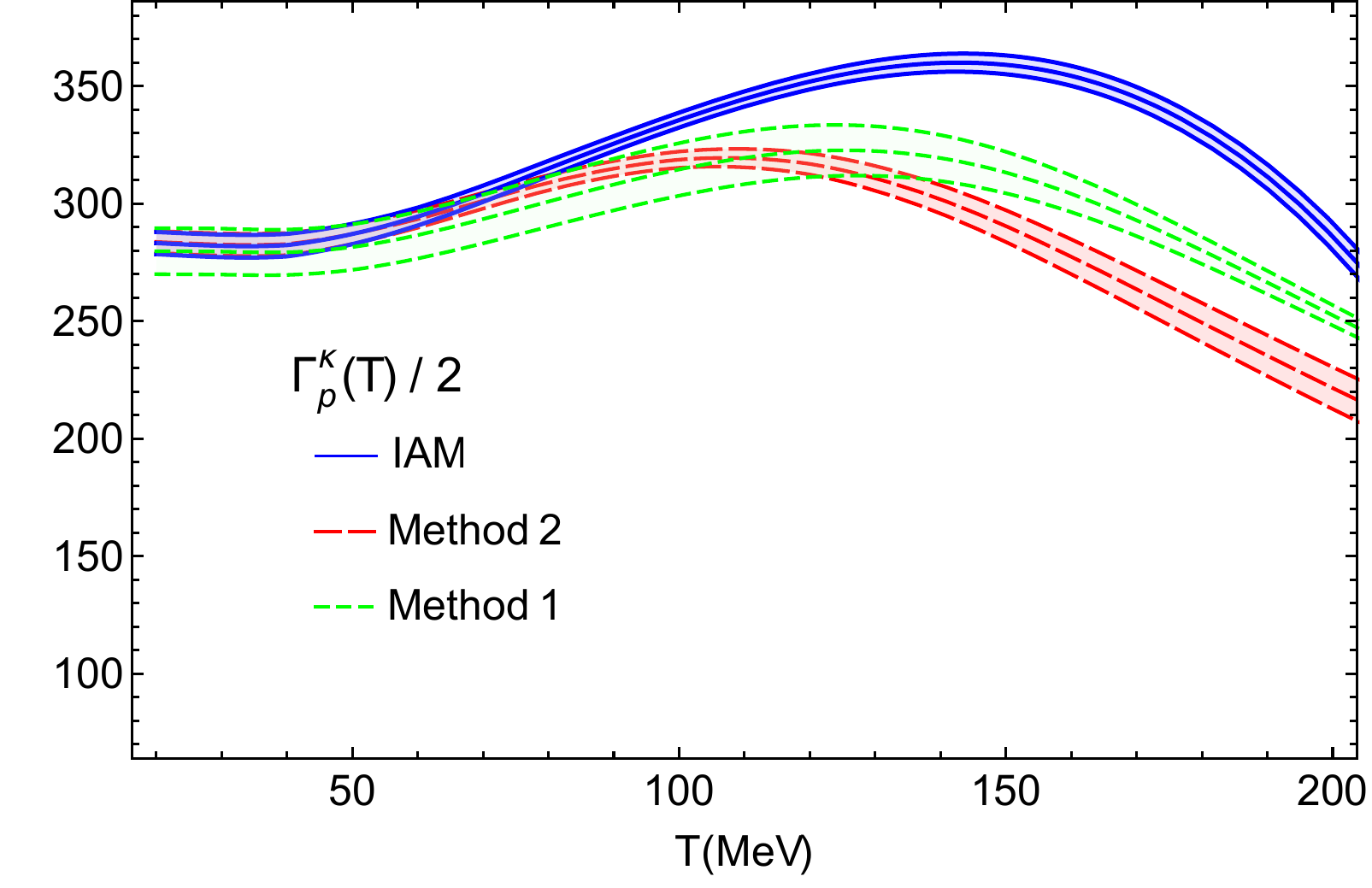}\\
\includegraphics[width=8cm]{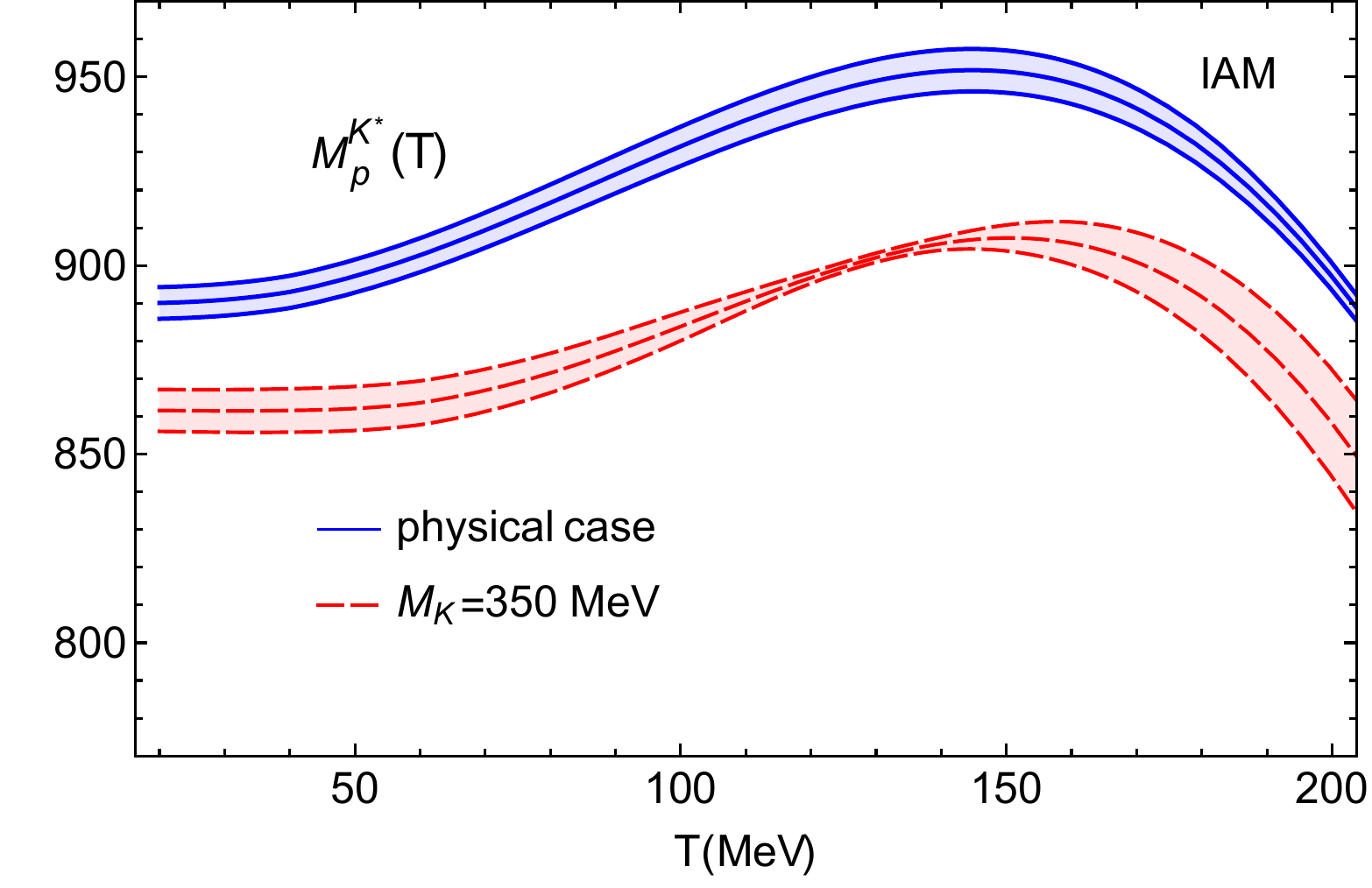}
\includegraphics[width=8cm]{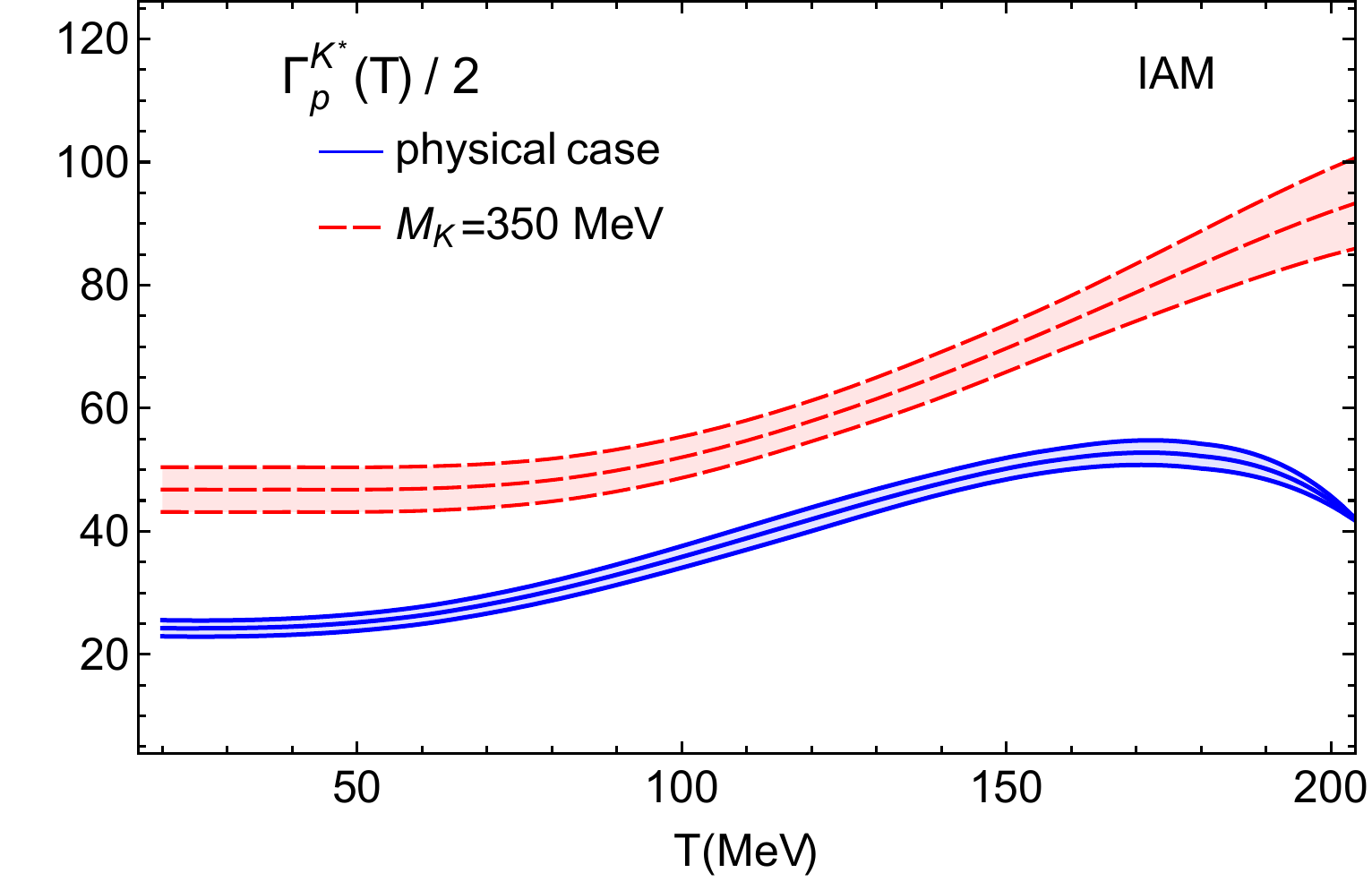}
\caption{Upper panels: Pole parameters $M_p$ (upper left panel) and $\Gamma_p$ (upper right panel) of the $K_0^* (700)$ ($\kappa$) at finite temperature calculated within the three different unitarization methods explained in the main text. Lower panels: $M_p$ (lower left panel) and $\Gamma_p$ (lower right panel) parameters of the $K^* (892)$ obtained within the IAM unitarization method at finite temperature. For the $K^* (892)$ we also show the result of reducing the kaon mass from its physical value to $M_K=350$ MeV.}
\label{fig:poles}
\end{figure}

 \begin{figure}[h]
 \centering
\includegraphics[width=12cm]{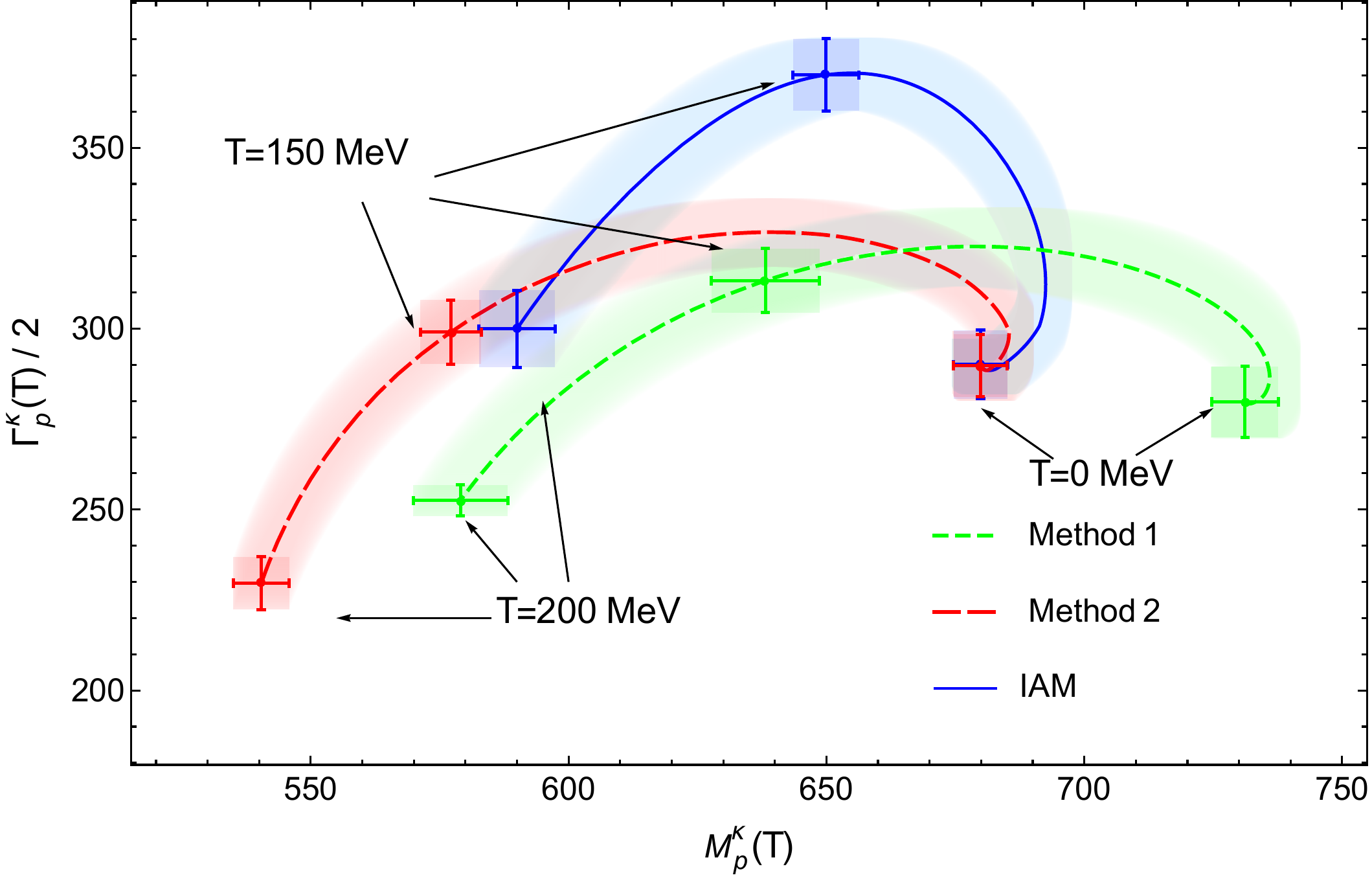}
\includegraphics[width=12cm]{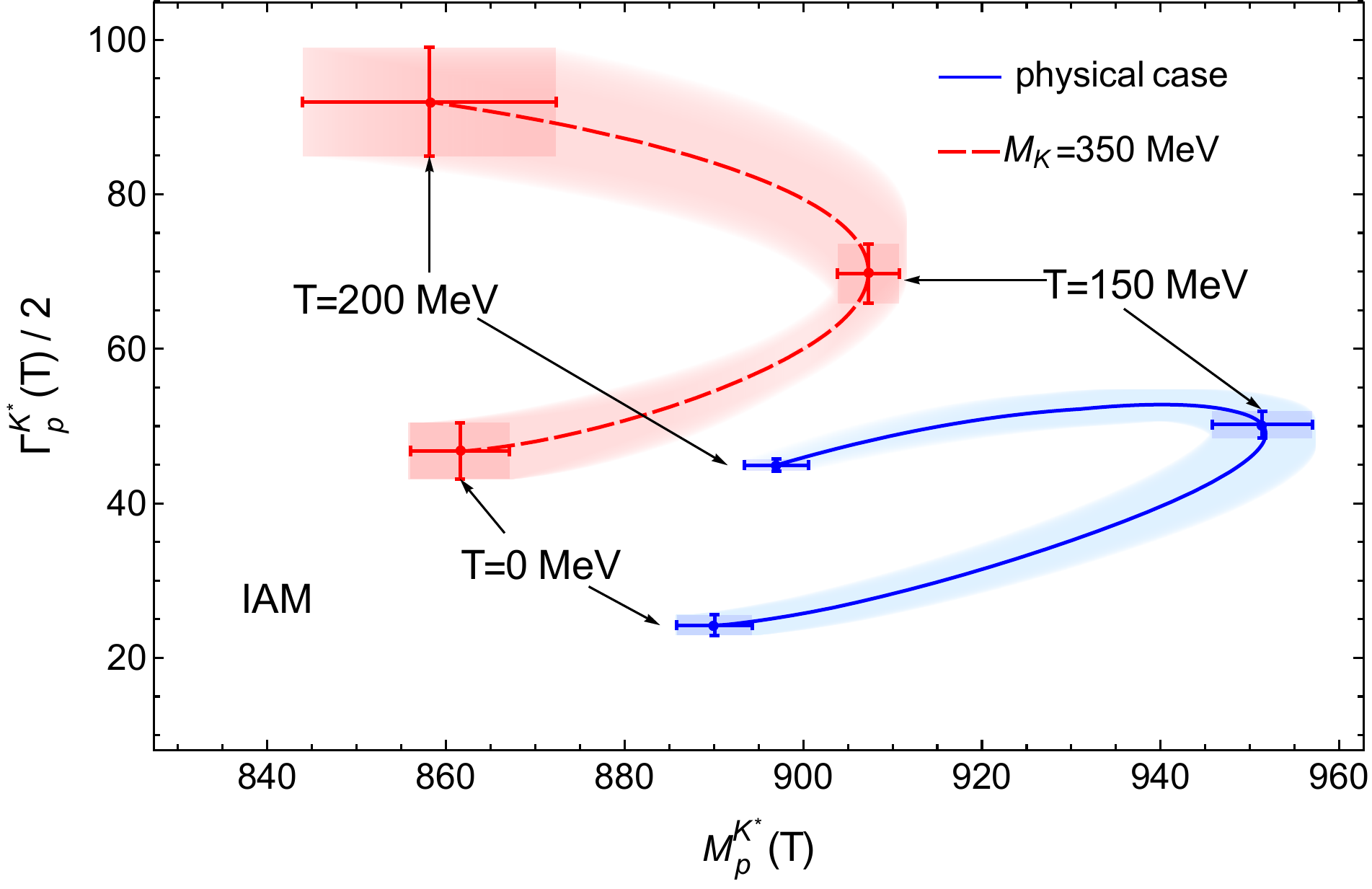}
\caption{Upper (lower) panel: the temperature dependence of the pole position of the $K_0^* (700)$ ($K^* (892)$) resonance when increasing
the temperature $T$ from 0 to 200 MeV.  The bands correspond to the LECs uncertainties. In both panels, we have also plotted the pole position at $T=0$, $150$, and $200$ MeV with their respective uncertainties. }
\label{fig:poleplane}
\end{figure}

In Fig.~\ref{fig:poles} and Fig.~\ref{fig:poleplane} we present our results for the $T$-dependence of the $K_0^* (700)$ and $K^* (892)$ poles. We have displayed in Fig.~\ref{fig:poles} the curves corresponding to the thermal dependence of the pole parameters $M_p$, $\Gamma_p$, while in  Fig.~\ref{fig:poleplane} we show the thermal pole trajectories, that is, we plot the width as a function of the mass when increasing the temperatures from $0$ to $200$ MeV. In the case of the $K_0^* (700)$, we compare our analysis with the results in~\cite{Gao:2019idb}, which we call method 1. The unitarized amplitude in method 1 includes only the contribution from the $J_{K\pi}(S)$ loop function in the fourth-order partial waves, i.e., it neglects the $T-U$ channel contributions and their modification of the analytic structure discussed in section~\ref{sec:anther}. In addition, it does not include any tadpole-like corrections.
A variant of this method was considered in~\cite{GomezNicola:2020qxo} (here method 2) by considering in the full one-loop ChPT amplitude at $T=0$ but including only the thermal modifications for the $J_{K\pi}(S)$ part. 

The results in Fig.~\ref{fig:poles} for the $K_0^* (700)$ confirm qualitatively the findings in~\cite{Gao:2019idb,GomezNicola:2020qxo}, although we observe sizable quantitative differences for $T>100$ MeV.
On the one hand, the real part of $\sqrt{s_p}$, i.e., $M_p(T)$, stays constant up to temperatures around $T\sim$ 75 MeV, from which it shows a decreasing behavior. On the other hand, $\Gamma_p(T)$ increases at low temperatures (roughly driven by the increase of thermal phase space in~\eqref{therps}) and decreases for $T$ closer to the transition, which can be understood as the regime where the reduction of phase space driven by the mass drop dominates over the increase given by \eqref{therps}. This behavior is very similar to that of the $f_0(500)$ pole obtained in~\cite{Dobado:2002xf,Ferreres-Sole:2018djq}, and in both cases  is fully consistent with the expected trends for the scalar susceptibility in terms of chiral and $U(1)_A$ restoration, as we will see in detail in section~\ref{sec:sus}. 

As for the $K^* (892)$, the results in Fig.~\ref{fig:poles} confirm a softer temperature dependence for the pole, consistently with the analysis in~\cite{Reichert:2022uha} based on heavy-ion data, which predict a softer medium dependence than for the $\rho (770)$, i.e., its corresponding $I=1$ partner in the vector octet. Actually, one of the main advantages of the IAM is that it encodes the correct quark-mass dependence inherited from ChPT; hence, with our present formalism, we can examine the beahvior towards the $SU(3)$ degeneration limit, $m_l/m_s\rightarrow 1$,
to check whether the $K^*(892)$ pole parameters become similar to those of the $\rho (770)$, for which $M_p$ varies slowly with $T$, but $\Gamma_p(T)$ increases considerably instead~\cite{Dobado:2002xf}. In our case, we confirm this behavior by reducing the kaon mass. For a kaon mass of $350$ MeV, we find that $\Gamma_p$ increases in all the considered temperature ranges. Moreover, as we can see in Fig.~\ref{fig:poles}, the width rises by $45$ MeV from $T=0$ to $T=200$ MeV, roughly doubling its value, while this gap is equal to $20$ MeV for the physical kaon mass. The variation of $M_p(T)$  on the contrary lies below the 10\% range in both cases.  

\begin{figure}[h]
 \centering
\includegraphics[width=10cm]{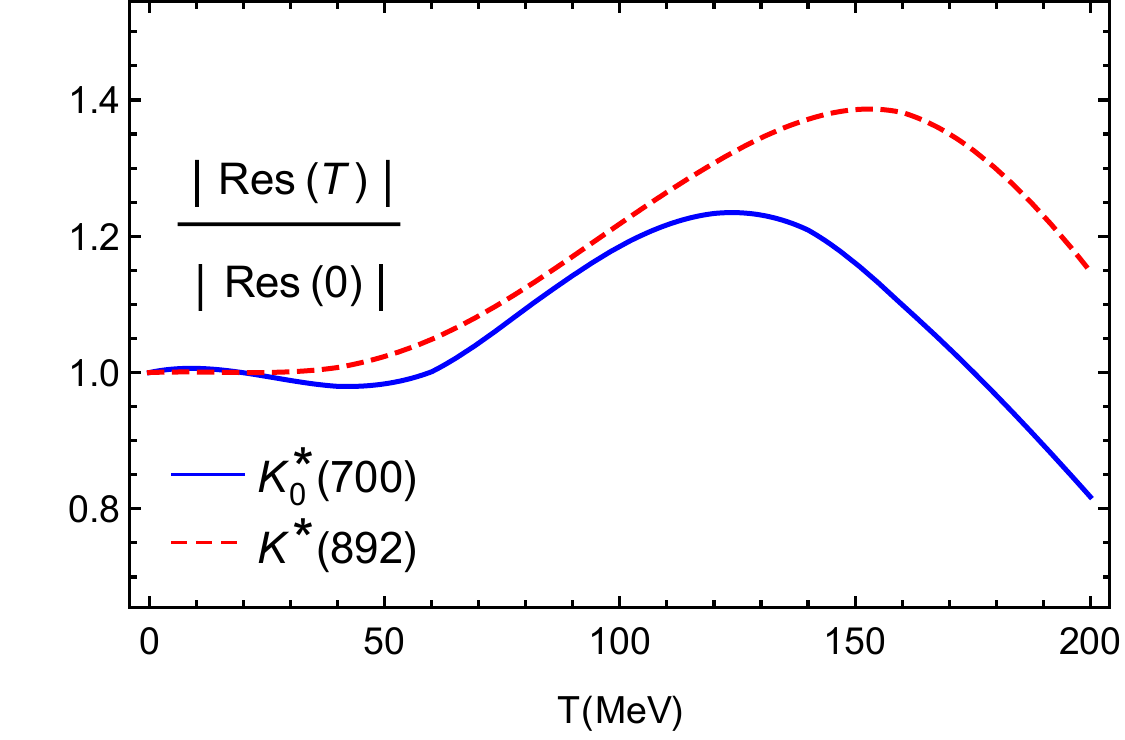}
\caption{Temperature dependence of the couplings of the $K_0^*$(700) and the $K^*(892)$ resonances to $K\pi$.}
\label{fig:residues}
\end{figure}

Finally, in Fig.~\ref{fig:residues}, we plot the thermal dependence of the modulus of the $K_0^*(700)$ and $K^*(892)$ pole residues,
obtained from the contour integral of the amplitudes in the second Riemann sheet around the pole position,
\begin{equation}
    \text{Res}\,(T)=\frac{1}{2\pi i}\oint \text{d}s\,t_{I J}(s,T)^{II}.
\end{equation}

In the case of elastic resonances, well-isolated from other singularity structures,  resonance residues can be related to the pole widths~\cite{Burkert:2022bqo}, which can be clearly observed by comparing Figs.~\ref{fig:residues} and~\ref{fig:poles}.
\footnote{In our convention, the thermal dependence of the width can be related to the product of the residue and thermal phase space for narrow resonances~\cite{Cabrera:2008tja}.}
Finally, note that the thermal dependence of both residues has a maximum in a temperature range between 120-170 MeV, with the $K^*(892)$ residue's maximum being slightly above the $K_0^*(700)$ one, confirming the correlation with the widths in Fig.~\ref{fig:poles}.

\section{The $I=1/2$ scalar susceptibility and consequences for chiral and $U(1)_A$ restoration}
\label{sec:sus}

The temperature dependence of the $K_0^* (700)$ pole discussed in section \ref{sec:unit} has important consequences regarding the restoration of both the chiral $SU(2)_V\times SU(2)_A\approx O(4)$ and $U(1)_A$ symmetries. The latter can be restored by medium effects related to instantons and its possible restoration close to the QCD transition has been the subject of several theoretical and lattice analyses in the literature 
\cite{Pisarski:1983ms,Shuryak:1993ee,Kapusta:1995ww,Cohen:1996ng,Lee:1996zy,Meggiolaro:2013swa,Buchoff:2013nra,Cossu:2013uua,Dick:2015twa,Nicola:2013vma,Brandt:2016daq,Tomiya:2016jwr,GomezNicola:2017bhm,Nicola:2018vug,GomezNicola:2020qxo,Nicola:2020iyl,Nicola:2020smo}. 

The connection with our present analysis comes from the following Ward Identities (WI) \cite{GomezNicola:2017bhm,Nicola:2018vug}

\begin{align}
\chi_P^K(T)=&\intT \left\langle{\cal T}K^a(x)K_a(0)\right\rangle=-\frac{\condl (T)+2\conds (T)}{m_l + m_s},
\label{WIK}\\
 \chi_S^\kappa(T)=&\intT \left\langle{\cal T}\kappa^a(x)\kappa_a(0)\right\rangle=\frac{\condl (T)-2\conds (T)}{m_s-m_l},
\label{WIkappa}
\end{align}
where $\chi_P^K$ and $\chi_S^\kappa$ are the pseudoscalar and scalar susceptibilities in the kaon and kappa channels, respectively, $\condl=\langle \bar u u + \bar d d\rangle$ and $\conds$ are the light- and strange-quark condensates, $m_{l,s}$ the light and strange quark masses, and 
\begin{equation}
  K^a=i\bar \psi\gamma^5\lambda^a \psi,\quad \kappa^a=\bar \psi\lambda^a \psi,\quad a=4,\cdots,7,\label{Kkbil}
\end{equation}
are the pseudoscalar and scalar $I=1/2$ quark bilinears, with $\psi$ the quark triplet, whose lightest states are  the kaon and $K_0^*(700)$ mesons, respectively. 

On the one hand, as explained in detail in~\cite{Nicola:2020smo}, the temperature dependence of the quark condensate combinations on the right-hand side of the WIs~\eqref{WIK}-\eqref{WIkappa} is consistent with the  degeneration of $K-\kappa$ susceptibilities at temperatures above the chiral transition $T_c$, signaling  $O(4)\times U(1)_A$ restoration in this channel~\cite{Nicola:2018vug}.
On the other hand, the WI~\eqref{WIkappa} predicts that $\chi_S^\kappa (T)$ must develop a peak above $T_c$. The behaviour of $\chi_S^\kappa$ below and above the peak is related to  $O(4)$ and  $O(4)\times U(1)_A$ restoration, respectively. In particular, it implies that when approaching the $m_l/m_s\rightarrow 1$ limit ($SU(3)$ degeneration) the peak should displace towards $T_c$ and increase its height, hence resembling the behavior of $\chi_S$, the $I=0$ scalar susceptibility, associated with the quantum numbers of the $f_0(500)$ resonance~\cite{Ferreres-Sole:2018djq}. Conversely, in the light chiral limit $m_l/m_s\rightarrow 0$, chiral symmetry restoration is enhanced, taking place at a lower $T_c$, and $K-\kappa$ degeneration takes place also at lower temperatures. As shown in \cite{GomezNicola:2020qxo}, this implies a more rapid growth of $\chi_S^\kappa (T)$ at the chiral transition region, i.e., below the  peak, which is confirmed by lattice results. On the contrary,  a flattening behavior is expected above the peak in the light chiral limit, consistently with a more efficient $K-\kappa$ degeneration~\cite{Nicola:2018vug,GomezNicola:2019myi,GomezNicola:2020qxo}.

Since the $K_0^*(700)$ or $\kappa$ is the lightest scalar state in the $I=1/2$ channel, we can expect it to provide the dominant contribution to $\chi_S^\kappa$~\cite{Nicola:2020smo}. Thus, as carried out also in \cite{GomezNicola:2020qxo}, our present analysis provides a way to compute $\chi_S^\kappa$ by saturating it with the thermal $K_0^*(700)$ pole, similarly to what was done in~\cite{Nicola:2013vma,Ferreres-Sole:2018djq} for the $\chi_S$ susceptibility in the $I=0$ channel, this time saturated with the $f_0(500)$ resonance.  For the $\kappa$ case, such an approach implies
\begin{equation}
\chi_S^{\kappa, U}(T)=A_\kappa \frac{M_\kappa^2 (0)}{M_\kappa^2 (T)},
\label{chisat}
\end{equation}
where $A_\kappa$ is fixed  to reproduce the perturbative ChPT result at $T=0$ calculated in~\cite{Nicola:2018vug}  and $M_\kappa^2 (T)=M_p^2(T)-\Gamma_p^2(T)/4$ is the real part of the  $K_0^*(700)$ self-energy at the pole, which in the present work is determined with the IAM, as discussed in section~\ref{sec:unit}.

\begin{figure}[h]
 \centering
\includegraphics[width=12cm]{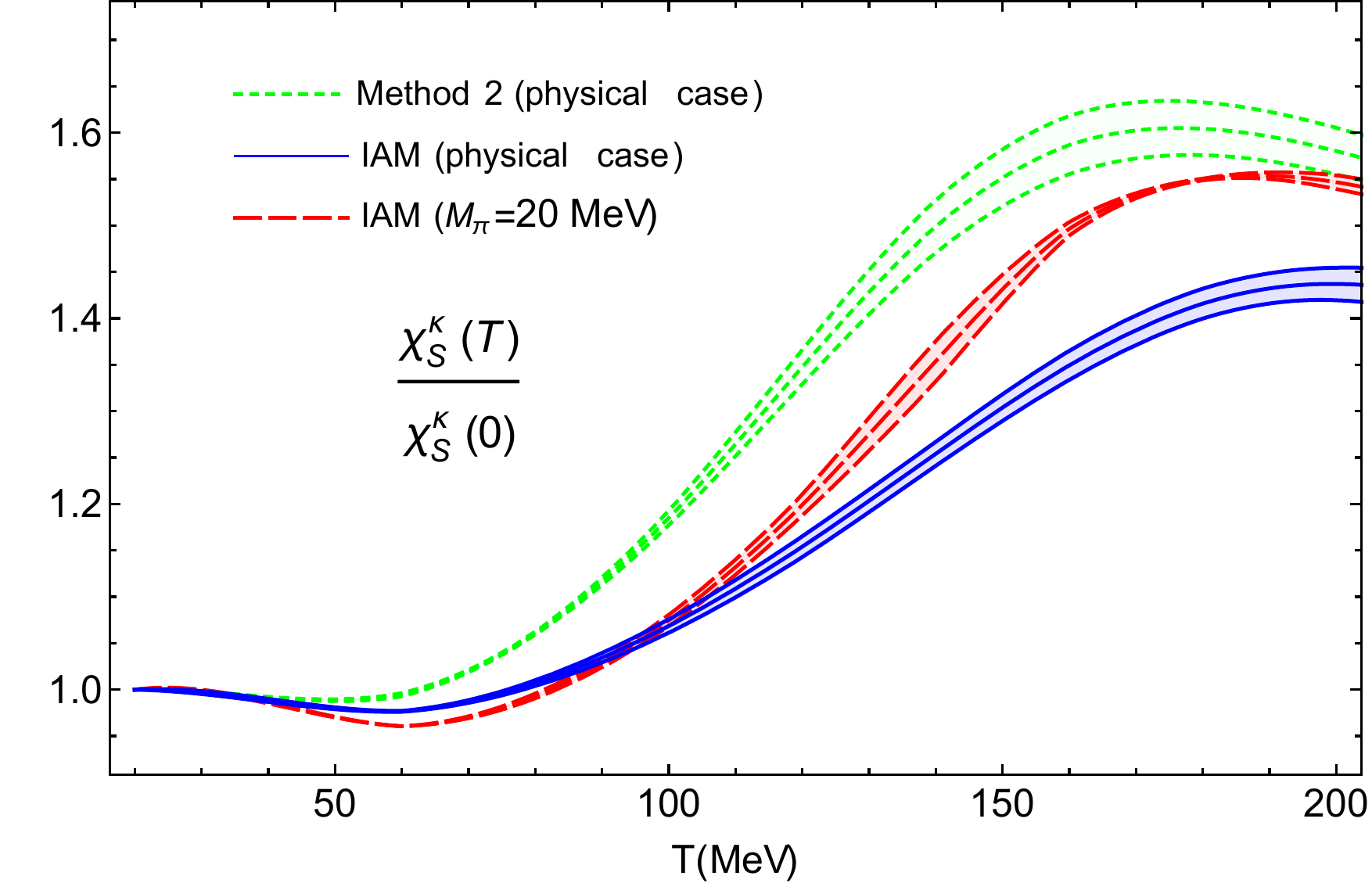}
\caption{Unitarized $\kappa$ susceptibility calculated using the IAM for physical masses and the light chiral limit, including the theoretical uncertainty bands of the LECs. We have also plotted the susceptibility obtained with the unitarization method 2.}
\label{fig:chikappachlim}
\end{figure}

\begin{figure}[h]
 \centering
\includegraphics[width=12cm]{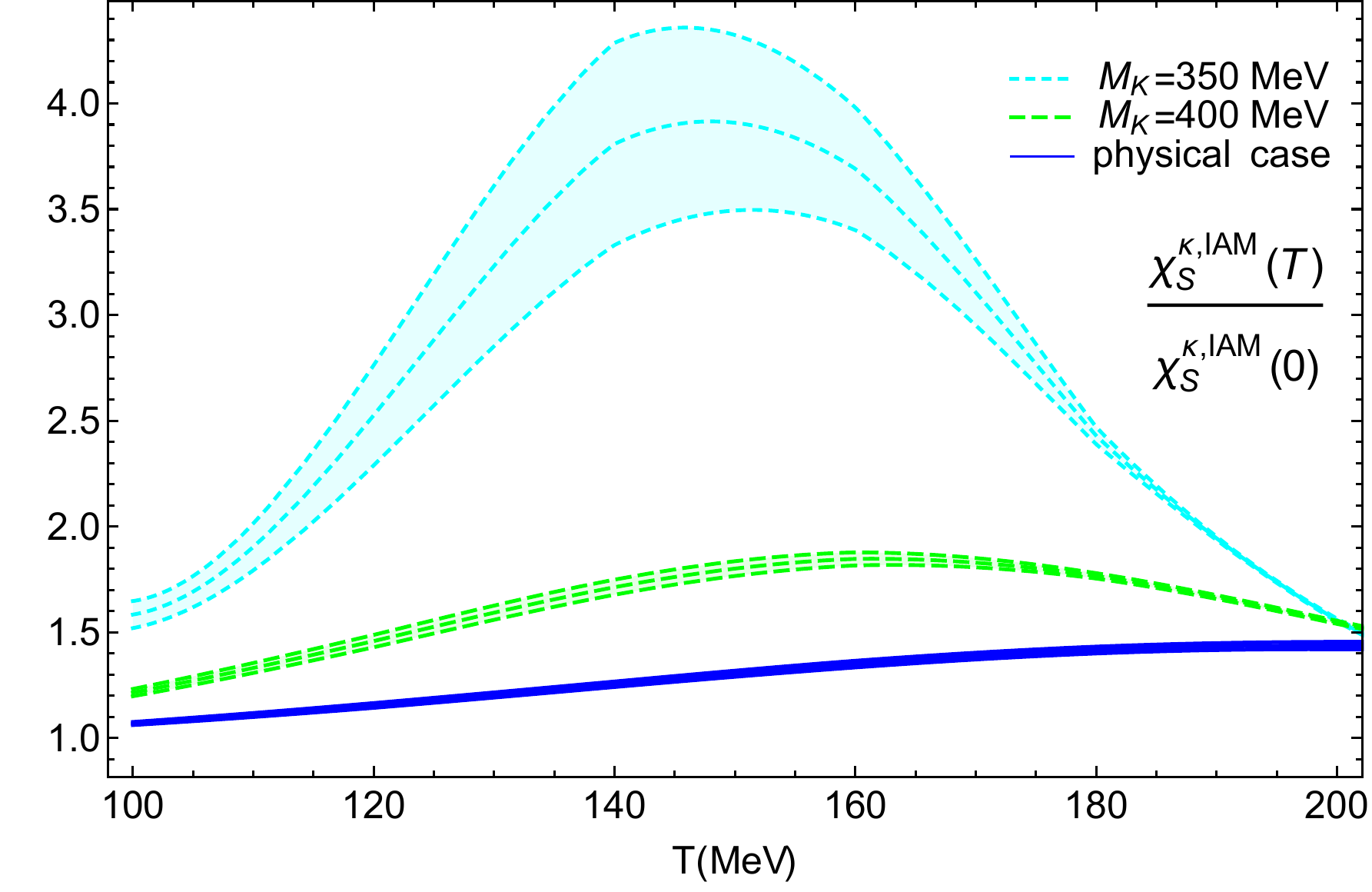}
\caption{Unitarized $\kappa$ susceptibility for physical masses and for $M_K$ closer to the SU(3) limit. The bands correspond to the LECs uncertainties.}
\label{fig:chikappamass}
\end{figure}

We show our results for $\chi_S^{\kappa, U}(T)$ in Figs.~\ref{fig:chikappachlim}~and~\ref{fig:chikappamass}. The peak of the $\kappa$ scalar susceptibility is clearly reproduced, as well as the expected behavior as the $m_l/m_s$ ratio is varied. In Fig.~\ref{fig:chikappachlim} we see that, near the light chiral limit of vanishing pion mass (with fixed $m_s$), the growth of the curve below the peak is more pronounced and compatible with a flattening above the peak. The evolution towards the $m_l/m_s\rightarrow 1$ limit is shown in Fig.~\ref{fig:chikappamass}, where we see the displacement of the peak towards $T_c$ and its growth, consistently with its degeneracy with $\chi_S$.  We also compare our complete unitarized ChPT calculation here with the method followed in~\cite{GomezNicola:2020qxo}.
The results are qualitatively similar, which supports their robustness.  In the chiral limit, the results in this work show more clearly the expected trend. 

\section{Conclusions}
\label{sec:conc} 

In this work, we have calculated the $\pi K$ elastic scattering amplitude at finite temperature $T$ in Chiral Perturbation Theory and we have obtained its unitarized version through the Inverse Amplitude Method, which allows one to generate the $K_0^*(700)$ and $K^*(890)$ poles. The thermal evolution of these states has been studied and related with a relevant application for the QCD phase diagram; namely, the connection of the $K_0^* (700)$ pole with chiral and $U(1)_A$ restoration through its role in the $I=1/2$ scalar susceptibility. 

The analytical structure of the amplitude shows interesting features at $T\neq 0$; apart from the standard $T=0$ cuts, including the unitarity cut above threshold, at $T\neq 0$ Landau cuts arise, which are related to physical processes taking place inside the thermal bath. The latter gives rise to an extended thermal unitarity relation, derived here for the first time for unequal-mass meson-meson scattering, such as the $\pi K$ case. 
 The unitarized thermal amplitude constructed here through the one-loop Inverse Amplitude Method satisfies exact unitarity, including the Landau-type processes, while matching the next-to-leading order ChPT series at low energies. 

By extracting the poles of the unitarized amplitude in the second Riemann sheet, we have obtained the thermal evolution of the $K_0^* (700)$ and the $K^{*}(892)$ resonances in the $I=1/2,\,J=0$ and $I=1/2,\,J=1$ channels, respectively. In addition, we have analyzed their behavior in the light chiral limit $m_l/m_s\rightarrow  0$ and when approaching the $SU(3)$ $m_l/m_s\rightarrow 1$ limit, both reachable within our theoretical framework. 

In the vector channel, the $K^{*}(892)$ pole position varies much more smoothly with temperature than its corresponding $I=1$ state in the vector meson nonet, i.e., the $\rho (770)$. This is in accordance with recent estimates based on heavy-ion data. We have examined the bevavior towards the $m_l/m_s\rightarrow 1$ limit, where the $SU(3)$ nonet members are meant to share physical properties. This analysis provides the expected results since the width of the $K^{*}(892)$ increases significantly with the temperature when the value of the kaon mass approaches the pion mass, thus resembling the behavior of the $\rho(770)$ at finite $T$ in the literature.

As for the $K_0^{*} (700)$, we have explored in detail the connection of the pole at finite $T$ with properties regarding the restoration of chiral and  $U(1)_A$ symmetries, relevant for the QCD phase diagram. In particular, through a saturation approach, the pole of the $K_0^{*} (700)$ can be related to the scalar susceptibility in the $I=1/2$ channel, which has been previously shown using WI to develop a peak above the QCD transition temperature $T_c$.  The susceptibility obtained confirms that behavior and satisfies the expected trend around that peak as $m_l/m_s$ is varied. On the one hand, as we approach the $m_l/m_s\rightarrow 1$  limit, the susceptibility tends to show the same behavior as for the $I=J=0$ channel one, saturated by the $f_0(500)$. In that case, chiral restoration dominates, the susceptibility peak increases and its center moves towards $T_c$. On the other hand, in the $m_l/m_s\rightarrow 0$ light chiral limit, the susceptibility slope increases below the peak, driven by the amplification of chiral restoration effects, while it tends to flatten above the peak due to a more efficient $K-\kappa$ partner degeneration.

We have also computed the residues of both resonances and found that their thermal dependence resembles the behavior of the corresponding pole widths, and in particular, it shows a maximum in the region around 120-170 MeV.

The results obtained here complement in a rigorous way previous analyses regarding finite-temperature hadronic properties of relevance for the QCD phase diagram, revealing new theoretical features. In particular, the extension of finite-temperature scattering to unequal masses developed here and the corresponding extended thermal unitarity, open many future lines for analysis of scattering and resonances within the light hadron multiplets.

 \begin{acknowledgments}
 Work partially supported by research contract  PID2019-106080GB-C21 (spanish ``Ministerio de Ciencia e Innovaci\'on"),  the European Union Horizon 2020 research and innovation program under grant agreement No 824093. JRE acknowledges support from the Swiss National Science Foundation, project No.\ PZ00P2\_174228 and the Ramón y Cajal program (RYC2019-027605-I) of the Spanish MINECO and, A. V-R from a fellowship of the UCM predoctoral program.
 \end{acknowledgments}
 
\appendix
 
 \section{Kinematics of  $\pi K$ elastic scattering}
 \label{app:kin}
 For the process $K\pi\rightarrow K\pi$ with external 4-momenta $p_K\,p_\pi  \rightarrow p'_K\,p'_\pi$, in the center of momentum (CM) frame where $\vec{S}\equiv \vec{p}_K+\vec{p}_\pi=\vec{p'}_K+\vec{p'}_\pi=\vec{0}$, we have in the physical region $s\geq (M_K+M_\pi)^2$, 
 
 \begin{eqnarray}
 \vert \vec{p}_K \vert^2&=&\vert \vec{p}_\pi \vert^2=\vert \vec{p'}_K \vert^2=\vert \vec{p'}_\pi \vert^2=p_{CM}^2(s;M_K,M_\pi),\nonumber
  \\
 E_{K}^2&=&E_{K}^{'\,2}=\frac{(s+\Delta_{K\pi})^2}{4s},\nonumber\\
 E_{\pi}^2&=&E_{\pi}^{'\,2}=\frac{(s-\Delta_{K\pi})^2}{4s},\nonumber\\
 T_0&\equiv& =E_{K}-E'_{K}=0, \quad \quad t(s,\theta)=-\vert \vec{T}\vert^2=2p_{CM}^2 (s;M_K,M_\pi)(\cos\theta-1),\label{TCM}\\
 U_0(s)&\equiv& =E_{K}-E'_{\pi}=\frac{\Delta_{K\pi}}{\sqrt{s}}, \quad \quad u(s,\theta)=U_0^2-\vert \vec{U}\vert^2=\frac{\Delta_{K\pi}^2}{s}-2p_{CM}^2 (s;M_K,M_\pi)(\cos\theta+1),
 \label{UCM}
 \end{eqnarray}
 where $s=S_0=(p_{K,0}+p_{\pi,0})^2$, $\Delta_{K\pi}=M_K^2-M_\pi^2>0$, $\cos\theta=\vec{p}_K\cdot \vec{p'}_K/p_{CM}^2$ is the scattering angle and 
 \begin{equation}
 p_{CM}^2(s;M_a,M_b)=\frac{\left[s-(M_a+M_b)^2\right]\left[s-(M_a-M_b)^2\right]}{4s}, \label{pcm}
 \end{equation}
 the  two-body phase space being given by $\sigma_{ab}(s)=\displaystyle \frac{2}{\sqrt{s}}p_{CM}(s;M_a,M_b)$. 
 
 \section{Loop integrals at finite temperature}
 \label{app:loop}
 
 \subsection{General expressions and relations}
 
 The Matsubara sums in the loop integrals \eqref{tadpole} and \eqref{Jint}   can be performed using standard complex contour techniques \cite{galekapustabook}. We get
 
 \begin{eqnarray}
 F_{\beta a}(T)&=&F_{\beta i}(T=0)-\frac{T^2}{2\pi^2}\int_{M_a/T}^\infty dy \sqrt{y^2-(M_a/T)^2} n_B(Ty), \nonumber\\
 J_k^{ab}(Q_0,\modQ;T)&=&J_k^{ab}(Q^2;T=0) +
\frac{1}{4\pi^2} \int_0^\infty dq q^2 \int_{-1}^1 dx   \frac{1}{4E_1 E_2}\times\nonumber\\
&& \left\{  
 \frac{1}{E_1+E_2+Q_0+i\epsilon}\left[ (-E_1)^k n_B(E_1)+(Q_0+E_2)^k n_B(E_2) \right]
 \right.\nonumber\\
 &+&\left. 
 \frac{1}{E_1+E_2-Q_0-i\epsilon}\left[ (E_1)^k n_B(E_1)+(Q_0-E_2)^k n_B(E_2) \right]
 \right.\nonumber\\
 &-&\left. 
 \frac{1}{E_1-E_2+Q_0+i\epsilon}\left[ (-E_1)^k n_B(E_1)-(Q_0-E_2)^k n_B(E_2) \right]
 \right.\nonumber\\
 &-&\left. 
 \frac{1}{E_1-E_2-Q_0-i\epsilon}\left[ (E_1)^k n_B(E_1)-(Q_0+E_2)^k n_B(E_2) \right]
 \right\},
 \label{generalJ}
 \end{eqnarray}
 where the analytic continuation in \eqref{AC}  has been carried out, $q\equiv \vert \vec{q}\, \vert$, $E_1^2=q^2+M_a^2$, $E_2^2=q^2+\vert \vec{Q} \vert^2+M_b^2-2\vert \vec{q}\, \vert \vert \vec{Q} \vert x$, $n_B(y)=(e^{y/T}-1)^{-1}$ is the Bose-Einstein distribution function and the $T=0$ expressions are  \cite{Gasser:1984gg}:
 \begin{eqnarray}
 F_{\beta a}(0)&=&-2M_i^2\lambda-\frac{M_a^2}{16\pi^2}\log\frac{M_a^2}{\mu^2},\nonumber\\
 J_0^{ab}(Q^2;T=0)&=&-2\lambda-\frac{1}{16\pi^2\Delta_{ab}}\left[M_a^2\log\frac{M_a^2}{\mu^2}-M_b^2\log\frac{M_b^2}{\mu^2}\right]+\frac{1}{32\pi^2}\left[2+\left(\frac{\Delta_{ab}}{Q^2}-\frac{\Sigma_{ab}}{\Delta_{ab}}\right)\log\frac{M_b^2}{M_a^2}\right.\nonumber\\
&-&\left.\frac{\nu_{ab}(Q^2)}{Q^2}\log\frac{\left[Q^2+\nu_{ab}(Q^2)\right]^2-\Delta_{ab}^2}{\left[Q^2-\nu_{ab}(Q^2)\right]^2-\Delta_{ab}^2} \right],
 \end{eqnarray}
 where
\begin{equation}
\lambda=\frac{\mu^{D-4}}{16\pi^2}\left[\frac{1}{D-4}-\frac{1}{2}\left(\log 4\pi - \gamma+1 \right)\right],
\end{equation}
with $\mu$ the dimensional regularization scale, $\gamma$ the Euler constant, $\Delta_{ab}=M_a^2-M_b^2$, $\Sigma_{ab}=M_a^2+M_b^2$, $\nu_{ab}^2(x)=\left[x-(M_a+M_b)^2\right]\left[x-(M_a-M_b)^2\right]$
and $J_{1,2} (Q^2;0)$ are related to $J_0 (Q^2;0)$ by the usual Veltman-Pasarino relations provided also in \cite{Gasser:1984gg}.  As mentioned above, those relations are no longer valid at nonzero $T$. Instead, at finite temperature, the following relations hold for the different combinations of loop integrals with momenta in the numerator (with $q_0=2\pi n T$, $Q_0=2\pi m T$ and Euclidean metric $(-,-,-,-)$):

	

\begin{equation}
	\sumint_q
	\dfrac{q_i}{\left(q^2-M_a^2\right)\left((q-Q)^2-M_b^2\right)}=-\dfrac{Q_i}{|\vec{Q}|^2}\left[Q_0J_1+\dfrac{(Q^2+\Delta_{ab})}{2}J_0+\dfrac{1}{2}\left(F_{\beta b}-F_{\beta a}\right)\right],
	\label{relJ1}
	\end{equation}
	
	\begin{equation}
	\sumint_q
	\dfrac{q_{\mu}Q^{\mu}}{\left(q^2-M_a^2\right)\left((q-Q)^2-M_{b}^2\right)}=\dfrac{(Q^2+\Delta_{ab})}{2}J_0+\dfrac{1}{2}\left(F_{\beta b}-F_{\beta a}\right),
	\end{equation}
	
	\begin{equation}
	\sumint_q
	\dfrac{q_{0}q_{i}}{\left(q^2-M_a^2\right)\left((q-Q)^2-M_{b}^2\right)}=-\dfrac{Q_i}{|\vec{Q}|^2}\left[Q_0J_2+\dfrac{(Q^2+\Delta_{ab})}{2}J_1+\dfrac{Q_0}{2}F_{\beta b}\right],
	\end{equation}
 
	\begin{equation}
	\sumint_q
	\dfrac{q_{i}q_{j}}{\left(q^2-M_a^2\right)\left((q-Q)^2-M_{b}^2\right)}=Q_iQ_jI_{ab}+g_{ij}\hat I_{ab}\,,
	\end{equation}

	\begin{equation}
	\sumint_q
	\dfrac{q_{0}(q\cdot q)}{\left(q^2-M_a^2\right)\left((q-Q)^2-M_{b}^2\right)}=Q_0F_{\beta b}+M_{a}^2J_1\,,
	\end{equation}
		
	\begin{equation}
	\sumint_q
	\dfrac{q_{i}(q\cdot q)}{\left(q^2-M_a^2\right)\left((q-Q)^2-M_{b}^2\right)}=Q_iF_{\beta b}-M_a^2\dfrac{Q_i}{|\vec{Q}|^2}\left[Q_0J_1+\dfrac{1}{2}(Q^2+\Delta_{ab})J_0+\dfrac{1}{2}\left(F_{\beta b}-F_{\beta a}\right)\right],
	\end{equation}
	
	\begin{equation}
	\sumint_q
	\dfrac{(q\cdot q)^2}{\left(q^2-M_a^2\right)\left((q-Q)^2-M_{b}^2\right)}=(Q^2+\Sigma_{ab})F_{\beta b}+M_a^4J_0\,,
	\end{equation}
	
	\begin{equation}
	\begin{split}
	I_{ab}&=\dfrac{1}{(D-2)|\vec{Q}|^4}\left\{\left[(D-1)Q_0^2+|\vec{Q}|^2\right]J_2+(D-1)Q_0(Q^2+\Delta_{ab})J_1\right.\\
	&\left.+\left[\dfrac{D-1}{4}(Q^2+\Delta_{ab})^2+M_{a}^2|\vec{Q}|^2\right]J_0+\left[\left(\dfrac{Q^2}{2}+Q_0^2\right)(D-1)+|\vec{Q}|^2\right]F_{\beta b}\right.\\
	&\left.+\dfrac{D-1}{4}(Q^2+\Delta_{ab})\left(F_{\beta b}-F_{\beta  a}\right)\right\},
	\end{split}
	\end{equation}
		
	\begin{equation}
	\begin{split}
	\hat I_{ab}&=\dfrac{1}{(D-2)|\vec{Q}|^2}\left\{-Q^2J_2+Q_0(Q^2+\Delta_{ab})J_1+\left[\dfrac{1}{4}(Q^2+\Delta_{ab})^2+M_{a}^2|\vec{Q}|^2\right]J_0\right.\\
	&\left.-\dfrac{Q^2}{2}F_{\beta b}+\dfrac{1}{4}(Q^2+\Delta_{ab})\left(F_{\beta b}-F_{\beta  a}\right)\right\}.
	\end{split}
	\end{equation}
	
	In the CM frame where $\vert \vec{Q} \vert=0$, the following additional simplifications take place:
	
\begin{equation}
J_{1}^{ab}(Q_0,0;T)= \frac{1}{2Q_0}\left[F_{\beta a}-F_{\beta b }- \left(-Q_0^2+\Delta_{ab}\right)J_{0}(Q_0,0;T)\right],
\label{relJ1CM}
\end{equation}

	\begin{equation}
	    J_2^{ab}(Q_0,0;T)= -\frac{F_{\beta b }}{2}+\frac{\left(-Q_0^2+\Delta_{ab}\right)}{4Q_0^2} \left[F_{\beta b}-F_{\beta a}+ \left(-Q_0^2+\Delta_{ab}\right)J_{0}(Q_0,0;T)\right],
	    \label{relJ2CM}
	\end{equation}

 \subsection{Analytical structure}
 \label{sec:Janstr}

\begin{figure}[h]
 \centering
\includegraphics[width=12cm]{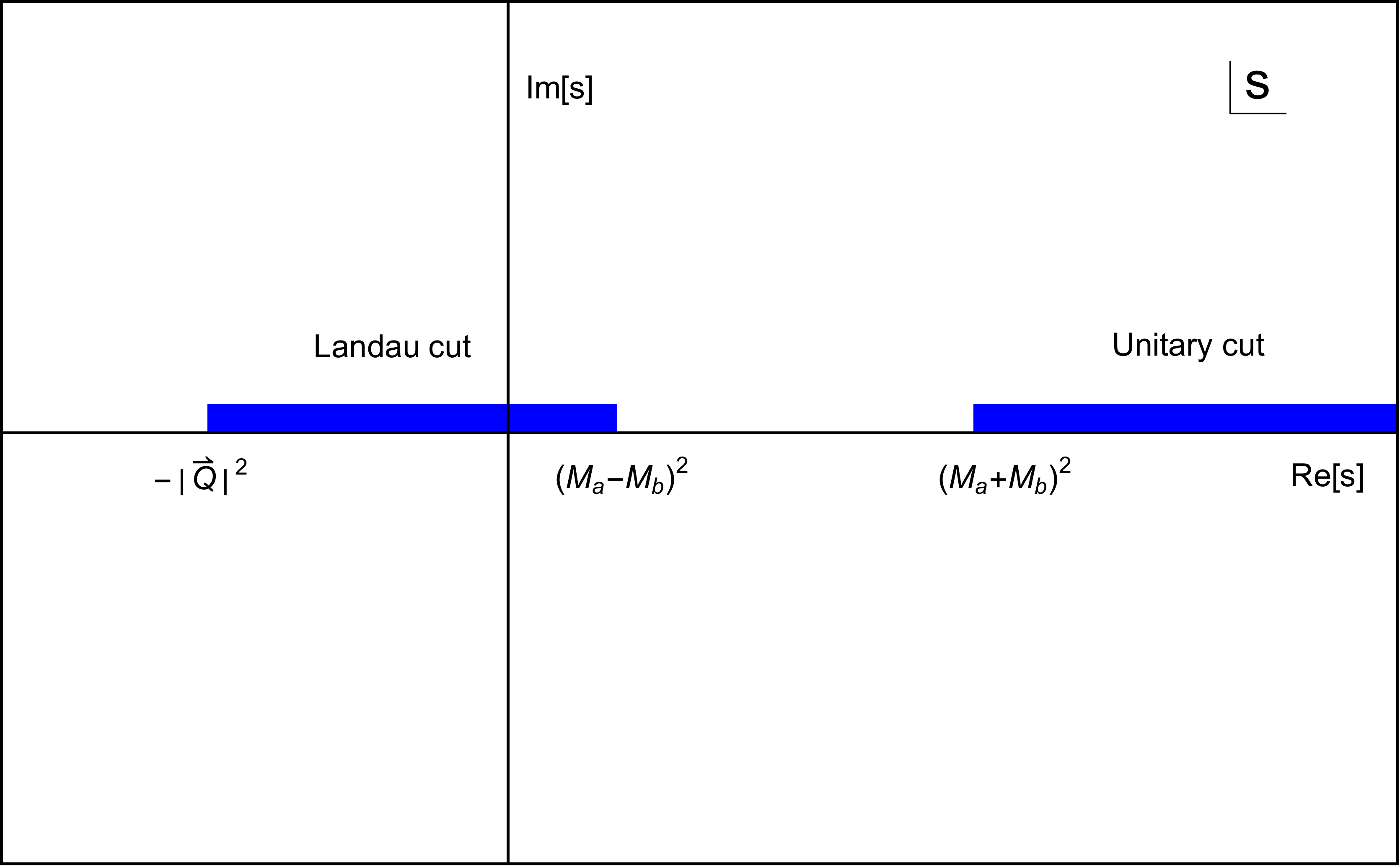}
\caption{Cuts of the $J_k^{ab}(Q_0,\modQ;T)$ integrals in \eqref{generalJ} in the $s=Q^2$ complex plane.}
\label{fig:Jcuts}
\end{figure}

Following the original analysis in~\cite{Weldon:1983jn}, one can use~\eqref{generalJ} to obtain the cuts in the real $Q_0$ axis for which the imaginary part of the $J_k$ integrals is nonzero. Note that it includes discontinuities related to physical processes inside the thermal bath. 
On the one hand, we have the standard unitary cut, which corresponds to the region of the integrand where $E_1+E_2=\vert Q_0 \vert$ (first two terms in~\eqref{generalJ}) requiring the condition $s\geq (M_a+M_b)^2$ with $s=Q_0^2-\vert \vec{Q} \vert^2$. This cut is the one giving rise to unitarity already at $T=0$, as discussed in section~\ref{sec:anther}. 
On the other hand, the third and fourth terms in \eqref{generalJ} give rise to the so-called Landau cuts, for $E_1-E_2=\vert Q_0 \vert$, which requires $-\vert \vec{Q} \vert^2 \leq s\leq (M_a-M_b)^2$ \cite{Ghosh:2010hap,Dasbook}. This cut is purely thermal, i.e., it vanishes at $T=0$. The analytic structure of the $J$ integrals is represented in Fig.~\ref{fig:Jcuts}. 

In section~\ref{sec:anther} we have discussed the interpretation of these cuts from physical processes taking place in the thermal bath. The contribution of those thermal processes becomes clearer by considering the imaginary part of the thermal integrals, namely, from~\eqref{generalJ} 
\begin{align}
\im J_k^{ab} (Q_0,\modQ;T) &= \frac{1}{2\pi^2} \int_0^\infty dq q^2 \int_{-1}^1 dx   \frac{E_1^k}{4E_1 E_2}\left\{ \left[1+n_B(E_1)+n_B(E_2)\right]\left[\delta(E_1+E_2-Q_0) + (-1)^{k+1}\delta(E_1+E_2+Q_0)  \right]
\right.\nonumber\\
&-\left.\left[n_B(E_1)-n_B(E_2)\right]\left[\delta(E_1-E_2-Q_0) + (-1)^{k+1} \delta(E_1-E_2+Q_0)\right]
\right\}.
\label{imJgen}
\end{align}

\subsection{Particular cases of interest for the scattering amplitude}
\label{sec:parcases}
 
 Here we provide useful expressions for the thermal loop integrals  corresponding to the scattering diagrams in Fig.~\ref{fig:diags} according to the center of mass kinematics provided in Appendix~\ref{app:kin}. 
 
\subsubsection {$S$-channel in CM frame}
 
 In this case, in the CM frame, we are interested in $Q_0^2=s\geq 0, \vert \vec{Q} \vert=0$ in \eqref{generalJ}. To derive the imaginary part in \eqref{imJgen}, recall that the solution to $\left[E_1(q)\pm E_2(q)\right]^2=s$ is given by $q=p_{\text CM}(s;M_a,M_b)$  in~\eqref{pcm} and $E_1=\vert s+\Delta_{ab} \vert/(2\sqrt{s})$, $E_2=\vert s-\Delta_{ab} \vert/(2\sqrt{s})$. In particular, note that $E_1=E_2$ for $M_a=M_b$. Therefore, 

 \begin{eqnarray}
 (E_1+E_2)^2=s &\Leftrightarrow & E_1=\frac{s+\Delta_{ab}}{2\sqrt{s}}, \quad E_2=\frac{s-\Delta_{ab}}{2\sqrt{s}} \qquad \left(s\geq (M_a+M_b)^2\right),
 \label{E1+E2s}
 \\
 (E_1-E_2)^2=s &\Leftrightarrow & E_1=\frac{s+\Delta_{ab}}{2\sqrt{s}}, \quad E_2=\frac{\Delta_{ab}-s}{2\sqrt{s}} \qquad \left(0\leq s\leq (M_a-M_b)^2\right),
 \label{E1-E2s}
 \end{eqnarray}
 where we are choosing the masses so that $M_a\geq M_b$, so that $\Delta_{ab}\geq 0$ and $E_1\geq E_2$ for $s\geq 0$. Note that in this frame, the lower limit for the Landau cut is at $s=0$, in accordance with Fig.~\ref{fig:Jcuts}. In this way, this cut vanishes for $M_a=M_b$ in the CM frame. 
 
 Therefore, solving for the corresponding $\delta$ functions in~\eqref{imJgen} we get 
 \begin{eqnarray}
 \im J_k^{ab} (Q_0,0;T)&=&\sgn(Q_0)^{k+1}\frac{\sigma_{ab}(s)}{16\pi}\left(\frac{s+\Delta_{ab}}{2\sqrt{s}}\right)^k \left\{  \left[ 1+n_B\left(\frac{s+\Delta_{ab}}{2\sqrt{s}}\right)+n_B\left(\frac{s-\Delta_{ab}}{2\sqrt{s}}\right)\right] \theta\left[s-(M_a+M_b)^2\right]
 \right.
 \nonumber\\
 &+& \left.
 \left[ n_B\left(\frac{\Delta_{ab}-s}{2\sqrt{s}}\right)-n_B\left(\frac{s+\Delta_{ab}}{2\sqrt{s}}\right)\right] \theta\left[(M_a-M_b)^2-s\right]\theta\left(s\right)
 \right\}, 
 \label{imJs}
 \end{eqnarray}
 with $s=Q_0^2$ and where we can readily identify the unitary and Landau contributions. The above structure has direct implications for thermal unitarity, as discussed in section \ref{sec:anther}. 
 
 For the extension of the thermal amplitude to the $s$ or $Q_0$ complex plane, we need an expression for the thermal integrals manifestly analytic in $Q_0+i\epsilon$. One can use  the relation
 \begin{equation}
 J_0^{ab}(s,0;T)-J_0^{ab}(s;T=0)=-\frac{1}{2\pi^2}\int_{M_1}^\infty dy \sqrt{y^2-M_1^2}\frac{s+\Delta_{ab}}{(s+\Delta_{ab})^2-4sy^2}n_B(y)+ \left(M_a\leftrightarrow M_b \right)
 \label{Js}
 \end{equation}
 where  $s=(Q_0+i\epsilon)^2=s+\sgn(Q_0)i\epsilon$ with $\epsilon\rightarrow 0^+$, so that the above expression allows one for the analytical continuation to the upper (lower) $s$ plane for $Q_0>0$ ($Q_0<0$). Since $J_0(s)$ is real in the real axis for $(M_a-M_b)^2\leq s \leq (M_a+M_b)^2$,  we can extend the analytical continuation from one half-plane to the other through Schwarz's reflection principle $J_0(\bar s)=\overline{J_0(s)}$. 
 Note that the corresponding analytic functions for $J_{1,2}(s)$ in the CM frame can be obtained from~\eqref{Js} through the relations~\eqref{relJ1CM}~and~\eqref{relJ2CM}. 
 
 The expressions in~\eqref{imJs}~and ~\eqref{Js} reduce to that in~\cite{GomezNicola:2002tn} for the case of equal masses $M_a=M_b$. In particular, as mentioned above, the Landau cut contribution in~\eqref{imJs} vanishes in that case. 
 
 \subsubsection {$T$-channel in CM frame}\label{app:t-channel}
 In this channel we have $Q_0=T_0=0$ and $\vert\vec{Q}\vert^2=\vert\vec{T}\vert^2=-t(s,\theta)=-2p_{\text CM}^2 (s;M_K,M_\pi)(\cos\theta-1)$. 
 
 First, from \eqref{imJgen}, we realize that the thermal imaginary part corresponding to the Landau cut in the $t$ variable vanishes in this channel since it is proportional to $\int \left[n_B(E_1)-n_B(E_2)\right]\delta(E_1-E_2)$, which corresponds to the lower end of the Landau cut in Fig.~\ref{fig:Jcuts}. As for the unitary cut, it requires $t\geq (M_a+M_b)^2$, so that from~\eqref{pcm} we conclude that it is present only for $s<0$, as in the $T=0$ case. 

 Therefore, in the physical region $s\geq (M_a+M_b)^2$, where $\vert \vec{T} \vert>0$ and $t<0$, the $J_k(0,\vert \vec{T} \vert=\sqrt{-t};T)$ are real. In particular,  from \eqref{generalJ}, we have for $t<0$
 \begin{eqnarray}
 J_0^{ab}(0,\sqrt{-t};T)-J_0^{ab}(t;T=0)&=&-\frac{1}{4\pi^2} \mbox{P} \int_{-1}^1 dx \int_0^\infty \text{d}q\, \frac{q^2}{E_1(q)} \frac{n_B\left[E_1(q)\right]}{t+2qx\sqrt{-t}+\Delta_{ab}}+\left(M_a\leftrightarrow M_b\right)\nonumber\\
 &=&\frac{1}{8\pi^2 \sqrt{-t}} \int_0^\infty \text{d}q\, \frac{q  n_B\left[E_1(q)\right]}{E_1(q)}\log\left\vert \frac{t+\Delta_{ab}-2q\sqrt{-t}}{t+\Delta_{ab}+2q\sqrt{-t}}\right\vert + \left(M_a\leftrightarrow M_b\right)
 \label{J0t},\nonumber\\
 J_1^{ab}(0,\sqrt{-t};T)&=&0,
 \label{J1t}\nonumber\\
 J_2^{ab}(0,\sqrt{-t};T)-J_2^{ab}(t;T=0)&=&
 \frac{1}{8\pi^2 \sqrt{-t}} \int_0^\infty \text{d}q\,  q E_1(q) n_B \left[E_1(q)\right] \log\left\vert \frac{t+\Delta_{ab}-2q\sqrt{-t}}{t+\Delta_{ab}+2q\sqrt{-t}}\right\vert + \left(M_a\leftrightarrow M_b\right),
 \label{J2t}
 \end{eqnarray}
 where P denotes Cauchy's principal value and where we have performed the change of variable $\vec{q}\rightarrow -\vec{q}+\vec{T}$ in the integrals containing $n_B(E_2)$ in \eqref{generalJ}. 

The above expressions generalize those obtained in \cite{GomezNicola:2002tn,Dobado:2002xf} for $M_a=M_b$. Actually, in the present work, only $J_{aa}$ integrals appear in the $T$ channel, so we refer to \cite{Dobado:2002xf} for the analytical continuation of those integrals to the complex $s-$ or $t-$plane beyond the real $t<0$ region.

 \subsubsection {$U$-channel in CM frame}

 Now we have $Q_0=U_0$ and $\vert\vec{Q}\vert^2=\vert\vec{U}\vert^2$, which explicit expressions in the CM frame are  given in~\eqref{UCM}. Thus, in principle, one has both  the Landau and unitary cut contributions in this  channel. Performing as before the change $\vec{q}\rightarrow -\vec{q}+\vec{U}$ in the integrals containing $n_B(E_2)$, for real values of $U_0$ and $\vert\vec{U}\vert>0$, we have from~\eqref{generalJ}
 \begin{eqnarray}
&&J_0^{ab}(U_0,\vert \vec{U} \vert;T)-J_0^{ab}(u;T=0)=
\nonumber\\
&&-\frac{1}{4\pi^2}  \int_0^\infty dq \frac{q^2 n_B \left[E_1(q)\right]}{E_1(q)} \int_{-1}^1 dx \frac{u+2q \vert \vec{U} \vert x+\Delta_{ab}}{(U_0+i\epsilon)^4-2(U_0+i\epsilon)^2\left(2q^2+\vert \vec{U} \vert^2-2q\vert \vec{U} \vert x+\Sigma_{ab}\right)+\left(\Delta_{ab}-\vert \vec{U} \vert^2+2q\vert \vec{U} \vert x\right)^2}
\nonumber\\
&&+ \left(M_a\leftrightarrow M_b\right),
\label{J0u}\\
&&J_1^{ab}(U_0,\vert \vec{U} \vert;T)-J_1^{ab}(u;T=0)=
\nonumber\\
&&\frac{U_0}{4\pi^2}  \int_0^\infty dq \hspace{0.1cm}q^2\int_{-1}^1 dx\left\lbrace -2 n_B \left[E_1(q)\right] \frac{E_1(q)}{(U_0+i\epsilon)^4-2(U_0+i\epsilon)^2\left(2q^2+\vert \vec{U} \vert^2-2q\vert \vec{U} \vert x+\Sigma_{ab}\right)+\left(\Delta_{ab}-\vert \vec{U} \vert^2+2q\vert \vec{U} \vert x\right)^2}\right.
\nonumber\\
&&+\left.\frac{n_B \left[\tilde{E}_2(q)\right]}{\tilde{E}_2(q)} \frac{2q^2-u-2q\vert \vec{U} \vert x+\Sigma_{ab}}{(U_0+i\epsilon)^4-2(U_0+i\epsilon)^2\left(2q^2+\vert \vec{U} \vert^2-2q\vert \vec{U} \vert x+\Sigma_{ab}\right)+\left(-\Delta_{ab}-\vert \vec{U} \vert^2+2q\vert \vec{U} \vert x\right)^2}\right\rbrace,\\
&&J_2^{ab}(U_0,\vert \vec{U} \vert;T)-J_2^{ab}(u;T=0)=
\nonumber\\
&&-\frac{1}{4\pi^2}  \int_0^\infty dq \hspace{0.1cm}q^2\int_{-1}^1 dx\left\lbrace n_B \left[E_1(q)\right] \frac{E_1(q)(u+2q \vert \vec{U} \vert x+\Delta_{ab})}{(U_0+i\epsilon)^4-2(U_0+i\epsilon)^2\left(2q^2+\vert \vec{U} \vert^2-2q\vert \vec{U} \vert x+\Sigma_{ab}\right)+\left(\Delta_{ab}-\vert \vec{U} \vert^2+2q\vert \vec{U} \vert x\right)^2}\right.
\nonumber\\
&&+\left.\frac{n_B \left[\tilde{E}_2(q)\right]}{\tilde{E}_2(q)} \frac{\left[(\tilde{E}_2^2(q)-U_0^2)^2-(\tilde{E}_2^2(q)+U_0^2)(E_1^2(q)+\vert \vec{U} \vert^2-2q\vert \vec{U} \vert x)\right]}{(U_0+i\epsilon)^4-2(U_0+i\epsilon)^2\left(2q^2+\vert \vec{U} \vert^2-2q\vert \vec{U} \vert x+\Sigma_{ab}\right)+\left(-\Delta_{ab}-\vert \vec{U} \vert^2+2q\vert \vec{U} \vert x\right)^2}\right\rbrace,
 \end{eqnarray}
 with $\tilde{E}_2^2(q)=q^2+M_b^2$.

The previous expressions can be evaluated numerically and  extended to complex $U_0$ and $\vert \vec{U} \vert$ values when needed. Actually, using these expressions, we have checked numerically that the general cut structure depicted in Fig.~\ref{fig:Jcuts} is fulfilled. 
 
 \section{Complete  expression for the thermal amplitude}\label{app:amplitude}

Here we list the expressions of the temperature-dependent corrections of the pion-kaon scattering amplitude to one loop in ChPT, with ${\cal T}_4^{F,J} (\Sr,\Tr,\Ur;T)={\cal T}_4^{F,J} (s,t,u;T=0)+\Delta{\cal T}_4^{F,J} (\Sr,\Tr,\Ur;T)$ and $\Delta f(T)\equiv f(T)-f(0)$,
 \begin{eqnarray}
 \nonumber
 \Delta{\cal T}_4^F (\Sr,\Tr,\Ur;T)&=&\dfrac{1}{f^4}\left\lbrace  -\dfrac{(\vec{k}\cdot\vec{T}) (\vec{p}\cdot\vec{T})}{2|\vec{T}|^4}(3 t + 4 |\vec{T}|^2)\left(\Delta F_{\beta\pi}+\dfrac{1}{2}\Delta F_{\beta K}\right)-\dfrac{(\vec{k}\cdot\vec{T})}{|\vec{T}|^2}\left(\dfrac{1}{18} (7 |\vec{T}|^2 - 18 p_0T_0)\Delta F_{\beta\pi}\right.\right.\\ \nonumber
 &&+\left.\left.\dfrac{1}{6}(|\vec{T}|^2 - 3 p_0T_0)\Delta F_{\beta K}\right)+\dfrac{(\vec{p}\cdot\vec{T})}{|\vec{T}|^2}\left(\dfrac{1}{18} (7 |\vec{T}|^2 + 18 k_0T_0)\Delta F_{\beta\pi}+\dfrac{1}{6}(|\vec{T}|^2 + 3 k_0T_0)\Delta F_{\beta K}\right)\right.\\ \nonumber
 &&+\left.\dfrac{(\vec{k}\cdot\vec{U}) (\vec{p}\cdot\vec{U})}{16|\vec{U}|^4}\left(-3 (3 (-M_K^2 + M_\pi^2 + u) + 8 |\vec{U}|^2)\Delta F_{\beta\pi}+9 (M_\eta^2 - 2 M_K^2 + M_\pi^2 - 2 u)\Delta F_{\beta K}\right.\right.\\ \nonumber
 &&-\left.\left.3(3 (M_\eta^2 - M_K^2 + u) + 8 |\vec{U}|^2)\Delta F_{\beta\eta}\right)+\dfrac{(\vec{k}\cdot\vec{U})}{12|\vec{U}|^2}\left(\dfrac{1}{2}(-4 |\vec{U}|^2 + 3 (M_K^2 + M_\pi^2 - u + 6p_0 U_0))\Delta F_{\beta\pi}\right.\right.\\\nonumber
 &&+\left.\left.(3 M_\eta^2 - M_K^2 - 2 M_\pi^2 - 3 u)\Delta F_{\beta K}-\dfrac{1}{2}(6 M_\eta^2 + M_K^2 - M_\pi^2 - 9 u + 12 |\vec{U}|^2 - 
  18 p_0 U_0)\Delta F_{\beta \eta} \right)\right.\\ \nonumber
  &&+\left.\dfrac{(\vec{p}\cdot\vec{U})}{12|\vec{U}|^2}\left( (7 |\vec{U}|^2 - 
  3 (M_K^2 - 2 M_\pi^2 + 2 u - 3 k_0U_0))\Delta F_{\beta\pi}-\dfrac{1}{2}(3 M_\eta^2 - 16 M_K^2 + 13 M_\pi^2 - 12 u)\Delta F_{\beta K}\right.\right.\\ \nonumber
  &&+\left.\left.\dfrac{1}{2}(3 M_\eta^2 - 10 M_K^2 + M_\pi^2 + 6 |\vec{U}|^2 + 
  18 k_0U_0)\Delta F_{\beta \eta}\right)+\dfrac{t(\vec{p}\cdot\vec{k})}{2|\vec{T}|^2}\left(\Delta F_{\beta\pi}+\dfrac{1}{2}\Delta F_{\beta K}\right)\right.\\ \nonumber
  &&+\left.\dfrac{3(\vec{p}\cdot\vec{k})}{16|\vec{U}|^2}\left(- (M_K^2 - M_\pi^2 - u)\Delta F_{\beta\pi}- (M_\eta^2 - 2 M_K^2 + M_\pi^2 - 2 u)\Delta F_{\beta K}+ (M_\eta^2 - M_K^2 + u)\Delta F_{\beta \eta}\right)\right.\\ \nonumber
  &&+\left.\dfrac{1}{72}\left[M_K^2 -47 M_\pi^2 +7 s - 21 t - 20 u  + 
 28 k_0 T_0 - 
 28 p_0 T_0 + 
 12 k_0 U_0 - 
 42 p_0 U_0\right]\Delta F_{\beta\pi}\right.\\ \nonumber
 &&-\left.\dfrac{1}{180}\left[15 M_\eta^2+8 M_K^2+33 M_\pi^2-14 s+16 t-14 u-30 k_0T_0+30 p_0T_0\right]\Delta F_{\beta K}\right.\\ 
 &&+\left.\dfrac{1}{360}\left[39 M_\eta^2 - 199 M_K^2 + 46 M_\pi^2 +21 s + 21 t - 114 u + 
 180 k_0U_0 - 
 90 p_0U_0\right]\Delta F_{\beta \eta}\right\rbrace.
 \label{TF}
 \end{eqnarray}
 
 \begin{eqnarray}
 \Delta{\cal T}^{J} (\Sr,\Tr,\Ur;T)&=&\dfrac{1}{f^4}\left\lbrace\dfrac{1}{4}(M_K^2 + M_\pi^2 - s)^2\Delta J_0^{K\pi}(S)-\dfrac{\vec{k}\cdot\vec{p}}{4|\vec{T}|^2}\left[(t^2 + 4 M_\pi^2 |\vec{T}|^2)\Delta J_0^{\pi\pi}(T)-4tT_0\Delta J_1^{\pi\pi}(T)\right.\right.\\\nonumber
 &&\left.\left. +4t\Delta J_2^{\pi\pi}(T)+\dfrac{1}{2}(t^2 + 4 M_K^2 |\vec{T}|^2)\Delta J_0^{KK}(T)-2tT_0\Delta J_1^{KK}(T)+2t\Delta J_2^{KK}(T)\right]\right.\\ \nonumber
&&\left.+6\dfrac{(\vec{k}\cdot\vec{T}) (\vec{p}\cdot\vec{T})}{|\vec{T}|^4}\left[\dfrac{1}{36}(27 t^2 + 36 (M_\pi^2 + t) |\vec{T}|^2 + 14 |\vec{T}|^4)\Delta J_0^{\pi\pi}(T)-(3 t + 2 |\vec{T}|^2) T_0\Delta J_1^{\pi\pi}(T)\right.\right.\\ \nonumber
 &&\left.\left.-(|\vec{T}|^2 - 3 T_0^2)\Delta J_2^{\pi\pi}(T)+\dfrac{1}{24}(9 t^2 + 4 (3 M_K^2 + 2 t) |\vec{T}|^2)\Delta J_0^{KK}(T)-\dfrac{1}{6}(9 t + 4 |\vec{T}|^2) T_0\Delta J_1^{KK}(T)\right.\right.\\ \nonumber
 &&\left.\left.-\dfrac{1}{2}(|\vec{T}|^2- 3 T_0^2)\Delta J_2^{KK}(T)\right]+\dfrac{(\vec{k}\cdot\vec{T})}{|\vec{T}|^2}\left[\dfrac{1}{36}\left(6 t (t - 4 p_0T_0) + |\vec{T}|^2 (3 M_\pi^2 + 2 t - 14 p_0T_0)\right)\Delta J_0^{\pi\pi}(T)\right.\right.\\ \nonumber
 &&\left.\left.+\dfrac{1}{3}\left(-t T_0 + p_0(3 t + 2 |\vec{T}|^2 - 4 T_0^2)\right)\Delta J_1^{\pi\pi}(T)-2p_0T_0\Delta J_2^{\pi\pi}(T)-\dfrac{t}{12}(3 t + 4 |\vec{T}|^2)\Delta J_0^{KK}(T)\right.\right.\\ \nonumber
 &&\left.\left.+\dfrac{1}{6}\left((3 t + 4 |\vec{T}|^2) p_0 + 3 t T_0\right)\Delta J_1^{KK}(T)-p_0T_0\Delta J_2^{KK}(T)-\dfrac{1}{12}M_\pi^2|\vec{T}|^2\Delta J_0^{\eta\eta}(T)\right]\right.
 \\ \nonumber
&&\left.+6\dfrac{(\vec{p}\cdot\vec{T})}{|\vec{T}|^2}\left[\dfrac{1}{18}(3 t (t - 2 p_0T_0) + |\vec{T}|^2 (5 t - 7 p_0T_0))\Delta J_0^{\pi\pi}(T)+\dfrac{1}{3}(-t T_0+ k_0 (3 t + 4 |\vec{T}|^2 - 2 T_0^2))\Delta J_1^{\pi\pi}(T)\right.\right.\\ \nonumber
 &&\left.\left. -2k_0T_0\Delta J_2^{\pi\pi}(T)-\dfrac{t}{12}(t + 4 k_0T_0)\Delta J_0^{KK}(T)+\dfrac{1}{6}(t T_0 + k_0 (3 t - 4 T_0^2))\Delta J_1^{KK}(T)-k_0T_0\Delta J_2^{KK}(T)\right]\right.
 \\ \nonumber
 &&\left.-\dfrac{3\vec{k}\cdot\vec{p}}{16|\vec{U}|^2}\left[\left((M_K^2 - M_\pi^2 + u)^2 + 4 M_K^2|\vec{U}|^2\right)\Delta J_0^{K\pi}(U)-4U_0(M_K^2 - M_\pi^2 + u)\Delta J_1^{K\pi}(U)+4u\Delta J_2^{K\pi}(U)\right.\right.\\ \nonumber
 &&\left.\left.+\left((-M_\eta^2 + M_K^2 + u)^2 + 4 M_K^2|\vec{U}|^2 \right)\Delta J_0^{K\eta}(U)+4 U_0(M_\eta^2 - M_K^2 - u)\Delta J_1^{K\eta}(U)+4u\Delta J_2^{K\eta}(U)\right]\right.\\ \nonumber
&&\left.+\dfrac{3(\vec{k}\cdot\vec{U}) (\vec{p}\cdot\vec{U})}{16|\vec{U}|^4}\left[(3 (M_K^2 - M_\pi^2 + u)^2 + 4 M_K^2|\vec{U}|^2)\Delta J_0^{K\pi}(U)-12 U_0(M_K^2 - M_\pi^2 + u)\Delta J_1^{K\pi}(U)\right.\right.\\ \nonumber
&&\left.\left.-4(|\vec{U}|^2-3U_0^2)\Delta J_2^{K\pi}(U)+\left(3 (-M_\eta^2 + M_K^2 + u)^2 + 4 M_K^2 |\vec{U}|^2\right)\Delta J_0^{K\eta}(U)\right.\right.\\ \nonumber
 &&\left.\left.-12U_0(-M_\eta^2 + M_K^2 + u)\Delta J_1^{K\eta}(U)-4(|\vec{U}|^2 - 3 
U_0^2)\Delta J_2^{K\eta}(U)\right]\right.
 \\ \nonumber
 &&\left.-\dfrac{(\vec{k}\cdot\vec{U})}{8|\vec{U}|^2}\bigg[ \left(-M_K^4 + (M_\pi^2 - u)^2\right)\Delta J_0^{K\pi}(U)-2 (3 (M_K^2 - M_\pi^2 + u) p_0 - (M_K^2 + M_\pi^2 - u) U_0)\Delta J_1^{K\pi}(U)\right.\\ \nonumber
 &&\left. +12 p_0 U_0\Delta J_2^{K\pi}(U)-\dfrac{1}{3}(6 M_\eta^2 + M_K^2 - M_\pi^2 - 9 u) (M_\eta^2 - M_K^2 - u)\Delta J_0^{K\eta}(U)\right.\\ \nonumber
 &&\left.+\dfrac{2}{3}(9 (M_\eta^2 - M_K^2 - u) p_0 + (-6 M_\eta^2 - M_K^2 + M_\pi^2 + 9 u)U_0)\Delta J_1^{K\eta}(U)+12p_0U_0\Delta J_2^{K\eta}(U)\bigg] \right.
 \\ \nonumber
 &&\left.-\dfrac{(\vec{p}\cdot\vec{U})}{4|\vec{U}|^2}\bigg[\left(M_K^4 - 3 M_K^2 (M_\pi^2 - u) + 2 (M_\pi^2 - u)^2\right)\Delta J_0^{K\pi}(U)\right.\\\nonumber
&&\left.-\left(3 (M_K^2 - M_\pi^2 + u) k_0 + 
  2 (M_K^2 - 2 M_\pi^2 + 2 u) U_0\right)\Delta J_1^{K\pi}(U)\right.\\ \nonumber
 &&\left.+ 6 k_0 U_0\Delta J_2^{K\pi}(U)+\dfrac{1}{6}(3 M_\eta^2 - 10 M_K^2 + M_\pi^2) (M_\eta^2 - M_K^2 - u)\Delta J_0^{K\eta}(U)\right.\\ \nonumber
 &&\left.+\dfrac{1}{3} \left(9 (M_\eta^2 - M_K^2 - u) k_0 + (3 M_\eta^2 - 10 M_K^2 + M_\pi^2) U_0\right)\Delta J_1^{K\eta}(U)+ 6 k_0U_0\Delta J_2^{K\eta}(U)\bigg] \right.
 \\ \nonumber
 &&\left.+ \dfrac{1}{4}(M_\pi^2 - u) (M_K^2 + M_\pi^2 - u)\Delta J_0^{K\pi}(U)-\dfrac{1}{4} \left(-(M_K^2 + M_\pi^2 - u) k_0 + 
 2 (M_K^2 - 2 M_\pi^2 + 2 u) p_0\right)\Delta J_1^{K\pi}(U)\right.
 \\ \nonumber
 &&\left.+\dfrac{3}{3}k_0 p_0\Delta J_2^{K\pi}(U)-\dfrac{1}{216}(3 M_\eta^2 - 10 M_K^2 + M_\pi^2) (6 M_\eta^2 + M_K^2 - M_\pi^2 - 9 u)\Delta J_0^{K\eta}(U)\right.
 \\ \nonumber
 &&\left.-\dfrac{1}{12}\left((6 M_\eta^2 + M_K^2 - M_\pi^2 - 9 u) k_0 - (3 M_\eta^2 - 10 M_K^2 + M_\pi^2) p_0\right)\Delta J_1^{K\eta}(U)-\dfrac{3}{2}\Delta J_2^{K\eta}(U)\right.
 \\ \nonumber
 &&\left.+\dfrac{t}{36}\left( (-3 M_\pi^2 + 4 t - 10 p_0T_0) + k_0T_0 (3 M_\pi^2 + 2 t - 14 p_0T_0)\right)\Delta J_0^{\pi\pi}(T)+\dfrac{1}{3}\left(t p_0 + k_0 (t + 6 p_0T_0)\right)\Delta J_1^{\pi\pi}(T)\right.
 \\ \nonumber
 &&\left.+2k_0p_0\Delta J_2^{\pi\pi}(T)+\dfrac{t}{12}(t + 4 k_0T_0)\Delta J_0^{KK}(T)-\dfrac{1}{6}\left(t p_0 + k_0 (3 t - 4 p_0 T_0)\right)\Delta J_1^{KK}(T)+k_0p_0\Delta J_2^{KK}(T)\right.\\ \nonumber
 &&\left.-\dfrac{1}{36}M_\pi^2 (3 M_\eta^2+M_\pi^2-3 t-3 k_0T_0)\Delta J_0^{\eta\eta}(T)
 \right\rbrace\,,
 \label{TJ}
 \end{eqnarray}
with $k_\mu=(p_K)_\mu$ and $p_\mu=(p_\pi)_\mu$.

\end{document}